\documentclass[prd,showpacs,superscriptaddress,nofootinbib,floatfix,showkeys,10pt]{revtex4-2}
\usepackage{bm}
\usepackage{times}
\usepackage{braket}
\usepackage{amsfonts,amssymb,stmaryrd,latexsym,amsmath}
\usepackage[usenames,dvipsnames]{color}
\usepackage{epsfig}
\usepackage{slashed}
\usepackage{hyperref}
\usepackage{graphics}
\usepackage{slashed}
\usepackage{orcidlink}
\usepackage{multirow}
\usepackage{caption}

\usepackage[export]{adjustbox}
\usepackage{array}

\usepackage{makecell}

\newcommand{\etal}{\textit{et al}. }
\renewcommand{\arraystretch}{2.0}

\allowdisplaybreaks[1]

\definecolor{nicered}{rgb}{0.7,0.1,0.1}
\definecolor{nicegreen}{rgb}{0.1,0.5,0.1}
\definecolor{emph}{rgb}{1,0,0}
\definecolor{doub}{rgb}{0.7,0.2,1.0}
\definecolor{navyblue}{RGB}{0, 110, 184}

\hypersetup{colorlinks,citecolor=nicegreen,linkcolor=nicered,urlcolor=navyblue}

\begin{document}
	
\title{Electromagnetic polarizabilities of the triplet hadrons in heavy hadron chiral perturbation theory} 

\author{Hao Dang\,\orcidlink{0000-0002-4320-2222}}
\email{haodang@stu.pku.edu.cn}
\affiliation{School of Physics, Peking University, Beijing 100871, China}

\author{Liang-Zhen Wen\,\orcidlink{0009-0006-8266-5840}}
\email{wenlzh\_hep-th@stu.pku.edu.cn}
\affiliation{School of Physics, Peking University, Beijing 100871, China}

\author{Yan-Ke Chen\,\orcidlink{0000-0002-9984-163X}}
\email{chenyanke@pku.edu.cn}
\affiliation{School of Physics and Center of High Energy Physics,
Peking University, Beijing 100871, China}

\author{Shi-Lin Zhu\,\orcidlink{0000-0002-4055-6906}}\email{zhusl@pku.edu.cn}
\affiliation{School of Physics and Center of High Energy Physics,
Peking University, Beijing 100871, China}

\begin{abstract}

We investigate the electromagnetic polarizabilities of singly heavy mesons and doubly heavy baryons within the framework of heavy hadron chiral perturbation theory up to $\mathcal{O}(p^3)$. We estimate the low-energy constants using the non-relativistic constituent quark model. A striking prediction of our study is the giant electric polarizabilities of the $D^*$ mesons: $\alpha_E(\bar{D}^{*0}) \approx 294 \times 10^{-4} \text{ fm}^3$ and $\alpha_E(D^{*-}) \approx 1.42-64.5 \text{i} \times 10^{-4} \text{ fm}^3$. 
These anomalously large values arise from the near-degenerate mass between $D^*$ and $D \pi$, which are orders of magnitude larger than those of their bottom counterparts. This kinematic coincidence induces a pronounced cusp structure in the chiral loops, reflecting the long-range dynamics of a pion cloud. For doubly heavy baryons, polarizabilities depend strongly on heavy-flavor composition: the $bcq$ system differs markedly from $ccq$ and $bbq$ due to mixing with scalar heavy-diquark states. Using heavy diquark–antiquark symmetry (HDAS), we unify the chiral dynamics of singly heavy mesons and doubly heavy baryons in the heavy-quark limit. The pion-loop contributions dominate the electromagnetic structure of heavy hadrons and provide essential benchmarks for future lattice QCD simulations.

\end{abstract}
 
\maketitle

\section{Introduction}

Revealing the internal structure of hadrons remains a central challenge in strong interaction physics, intimately tied to the non-perturbative dynamics of Quantum Chromodynamics (QCD) in the low-energy regime. Effective field theory (EFT), introduced by Weinberg through the construction of the most general Lagrangian consistent with the symmetries of QCD~\cite{Weinberg:1978kz}, provides a systematic and model-independent approach to this regime. Chiral perturbation theory ($\chi$PT), based on the approximate chiral symmetry of QCD, has achieved remarkable success in describing the low-energy dynamics of Goldstone bosons. However, for systems containing heavy hadrons, the presence of a large mass scale persisting in the chiral limit complicates standard chiral power counting. To address this, Heavy Hadron Chiral Perturbation Theory (HH$\chi$PT) was proposed to extend $\chi$PT to the heavy hadron sector~\cite{Jenkins:1990jv,Bernard:1992qa,Hemmert:1997ye}. By decomposing the heavy hadron field into ``light'' and ``heavy'' components and integrating out the heavy degrees of freedom, HH$\chi$PT restores a consistent power counting through an expansion in powers of the momentum (mass) of the pseudoscalar meson and the residual momentum of the heavy hadrons.

Electromagnetic observables provide clean probes of hadron structure due to the well-controlled nature of QED. The electromagnetic properties of heavy hadrons—such as electromagnetic form factors, radiative decays, and magnetic moments—have been extensively investigated using various theoretical approaches. These include constituent quark models (including bag models and potential models)~\cite{PhysRevD.56.348,PhysRevD.60.094002,PhysRevD.63.034005,Kumar_2005,PhysRevD.73.094013,PhysRevD.81.073001,Majethiya:2011ry,PhysRevD.96.116016,PhysRevD.104.053002,PhysRevD.36.2074,PhysRevD.53.1349,PhysRevD.75.073016,PhysRevD.94.113011,Ivanov:1994ji,PhysRevD.22.773,Simonis:2018rld,Bernotas:2012nz,PhysRevD.87.074016,PhysRevD.104.114011,PhysRevD.64.094007,Ebert:2002xz}, QCD sum rules and light-cone sum rules~\cite{AGAMALIEV201738,doi:10.1142/S021773231250054X,Aliev:1994nq,Dosch:1995kw,Zhu:1996qy,Zhu:1997as,Zhu:1998ih}, chiral soliton models~\cite{Scholl:2003ip,Oh:1991ws,Oh:1995eu,Kim:2021xpp}, Dyson-Schwinger equations~\cite{Xu:2024fun}, chiral effective field theory~\cite{PhysRevD.47.1030,PhysRevD.50.3295,Savage:1994wa,PhysRevD.71.054504,PhysRevD.92.054017,PhysRevD.96.076011,Meng:2017dni,Li:2017pxa,Wang:2019mhm,PhysRevD.99.034021,PhysRevD.98.054026}, and lattice QCD simulations~\cite{Becirevic:2009xp,Can:2013tna,Bahtiyar:2015sga,PhysRevD.92.114515,BAHTIYAR2017121,PhysRevD.98.114505,Can:2021ehb}. Electromagnetic polarizabilities, denoted as $\alpha_{E}$ and $\beta_{M}$, are fundamental observables that characterize the deformation of charge and magnetization distributions in response to external electromagnetic fields~\cite{Holstein:2013kia}. They are defined via the second-order effective Hamiltonian:
\begin{equation}\label{eq:effective_H}
	H^{(2)} = -\frac{1}{2}4\pi\alpha_E \vec{E}^2 - \frac{1}{2}4\pi\beta_M \vec{H}^2.
\end{equation} 
Their expressions for the particle state $|i\rangle$ can be derived from the second-order energy shift in perturbation theory~\cite{ERICSON1973604}:
\begin{equation}\label{eq:classcial_polar}
    \alpha_E = 2\alpha_{\rm em}\sum_{n \neq i} \frac{|\langle n|D_z|i\rangle|^2}{E_n - E_i}, \quad \beta_M = 2\alpha_{\rm em}\sum_{n \neq i} \frac{|\langle n|M_z|i\rangle|^2}{E_n - E_i},
\end{equation}
where we retain only the leading terms in the long-wavelength limit. $D_z$ and $M_z$ are the $z$-components of the electric and magnetic dipole operators, respectively. Eqs.~\eqref{eq:effective_H} and \eqref{eq:classcial_polar} indicate that the electromagnetic polarizabilities are dominated by contributions from near-degenerate states due to the small denominators.

In the chiral limit, the massless Goldstone bosons induce large pion-cloud fluctuations, leading to divergent contributions to polarizabilities. Despite explicit chiral symmetry breaking, which gives mass to the Goldstone bosons, chiral corrections arising from the interaction between photons and the pion-cloud remain substantial. HH\(\chi\)PT provides a robust framework for quantifying these effects. Within heavy-baryon chiral perturbation theory, Bernard \etal demonstrated the crucial role of $N\pi$ loops in nucleon polarizabilities, with results consistent with experimental data and lattice QCD~\cite{PhysRevLett.67.1515,Bernard:1991ru,Bernard:1993bg,Bernard:1993ry}. The calculations for singly heavy baryons also show that these long-range chiral corrections provide a significant contribution to electromagnetic polarizabilities~\cite{Chen:2024xks,Wen:2025xed},  which highlights the chiral dynamics playing an important role in non-perturbative effects.

Building on this foundation, we use the HH$\chi$PT formalism to systematically investigate the electromagnetic polarizabilities of hadrons in the flavor triplet representation, namely singly heavy mesons (e.g., $D, B$) and doubly heavy baryons (e.g., $\Xi_{cc}, \Omega_{bb}$). These systems provide a unique theoretical laboratory for probing non-perturbative QCD dynamics. In the heavy quark limit, the heavy quark or a compact heavy diquark acts as a static color source and decouples from the low-energy dynamics. This dynamical equivalence, known as the Heavy Diquark--Antiquark Symmetry (HDAS), establishes a deep connection between singly heavy mesons and doubly heavy baryons~\cite{SAVAGE1990177}. As a consequence, the response to external electromagnetic fields is governed predominantly by the light degrees of freedom, allowing us to isolate and quantify chiral dynamics within a unified heavy-flavor framework.

In this work, we calculate the electromagnetic polarizabilities of singly heavy mesons and doubly heavy baryons in HH\(\chi\)PT up to \(\mathcal{O}(p^3)\) and estimate the low-energy constants (LECs) using the approach in our previous work~\cite{Chen:2024xks,Wen:2025xed}. Although direct experimental measurements of these polarizabilities remain challenging due to the short lifetimes of heavy hadrons, our results provide well-defined theoretical benchmarks and offer valuable guidance for future lattice QCD simulations.

The remainder of this paper is organized as follows. In Sec.~\ref{sec:theoretical_framework}, we outline the general theoretical framework, including the derivation of electromagnetic polarizabilities from the spin-averaged forward Compton scattering amplitude and the construction of the relevant HH$\chi$PT effective Lagrangians. Section~\ref{sec:meson} presents analytical calculations and numerical results for singly heavy mesons, followed by the corresponding analysis for doubly heavy baryons in Sec.~\ref{sec:baryon}. 
Finally, Sec.~\ref{sec:DISCUSSION_AND_CONCULSION} summarizes our findings. The detailed loop integrals and explicit expressions for the form factors are collected in the Appendices~\ref{appendix:loop_integrals} and~\ref{appendix:expression_of_EMP}.

\section{theoretical framework}\label{sec:theoretical_framework}

\subsection{Definition of Polarizabilities via Spin-averaged forward Compton Scattering}\label{subsec:spin-averaged_forward_Compton_ tensor}
For a hadron with spin $s$, the electromagnetic polarizabilities are extracted through the spin-averaged forward Compton scattering tensor $\Theta^s_{\mu\nu}$~\cite{PhysRevLett.67.1515,Bernard:1991ru}:
\begin{equation}
   \Theta^s_{\mu\nu}=\frac{e^2}{2s+1}\sum_s\bar{\mathcal{U}}^s_{\alpha_1\alpha_2\dots}(p)T^{\alpha_1\alpha_2\dots,\beta_1\beta_2\dots}_{\mu\nu}(p,k)\mathcal{U}^s_{\beta_1\beta_2\dots}(p),
\end{equation}
where $k$ and $p$ denote the momentum of the photon and hadron, respectively. Here, $\mathcal{U}^s(p)$ represents the generalized wave function in momentum space. Its definition depends on the spin statistics of the hadron. For fermions ($s$ is a half-integer), $\mathcal{U}$ corresponds to the standard Dirac spinor $u(p)$, and the bar notation denotes the Dirac conjugate, $\overline{\mathcal{U}} \equiv u^\dagger \gamma^0$. For bosons ($s$ is an integer), $\mathcal{U}$ represents the polarization tensor (specifically, the polarization vector $\epsilon_\mu$ for $s=1$ and unity for $s=0$), while the bar notation is defined as the Hermitian conjugate, $\overline{\mathcal{U}} \equiv \mathcal{U}^\dagger$. The term 
\begin{equation}
    \bar{\mathcal{U}}^s_{\alpha_1\alpha_2\dots}(p)T^{\alpha_1\alpha_2\dots,\beta_1\beta_2\dots}_{\mu\nu}(p,k)\mathcal{U}^s_{\beta_1\beta_2\dots}(p)=\int d^4xe^{ik\cdot x}\left\langle\psi^s(p)\left|T\left[J_\mu^{\mathrm{em}}(x)J_\nu^{\mathrm{em}}(0)\right]\right|\psi^s(p)\right\rangle
\end{equation}
is the Fourier-transformed matrix element of two time-ordered electromagnetic currents for hadrons $\psi^s$. Within the HH$\chi$PT framework, the heavy hadron field $\psi(x)$ with mass $M$ is decomposed into ``light'' and ``heavy'' components, $\mathcal{B}(x)$ and $\mathcal{H}(x)$, as follows:
\begin{equation}
\psi(x) = e^{-iM v\cdot x} \left( \mathcal{B}(x) + \mathcal{H}(x) \right), \quad \text{with } \mathcal{B} = e^{iM v\cdot x} \frac{1+\slashed{v}}{2}\psi, \quad \mathcal{H} = e^{iM v\cdot x} \frac{1-\slashed{v}}{2}\psi,
\end{equation}
where $v^{\mu}=(1,\vec{0})$ represents the static velocity. The heavy component $\mathcal{H}(x)$ is then integrated out in the effective Lagrangians. In this formalism, the Dirac matrices are reduced to~\cite{Meng:2022ozq}
\begin{equation}
    \gamma_5\to 0, \quad \gamma_\mu\to v_\mu,\quad \gamma_\mu \gamma_5 \to 2S_\mu,\quad \sigma_{\mu\nu}\to -2\varepsilon_{\mu\nu\rho\sigma} v^\rho S^\sigma,
\end{equation}
where $S^\mu=\frac{i}{2}\gamma_5\sigma^{\mu\nu}v_\nu$ is the Pauli-Lubanski vector. 

The explicit expressions for $\Theta_{\mu\nu}^{s}$ after spin summation are provided in Table~\ref{tab:Theta_tensor}. Based on Lorentz covariance, one can decompose $\Theta^s_{\mu\nu}$ as~~\cite{Bernard:1993bg,Bernard:1993ry}
\begin{equation}
    \Theta^s_{\mu\nu}=U(\omega)g_{\mu\nu}+V(\omega)k_\mu k_\nu+W(\omega)\left(k_\mu v_\nu+v_\mu k_\nu\right)+X(\omega)v_\mu v_\nu,
\end{equation}
where $\omega=v\cdot k$ is the energy of the photon. Working in the Coulomb gauge ($\epsilon \cdot v = 0$), the contraction of $\Theta_{\mu\nu}^{s}$ with the photon polarization vectors simplifies to:
\begin{equation}\label{eq:lorentzexp}
	\epsilon^{\prime\mu}\Theta^s_{\mu\nu}\epsilon^\nu=(\epsilon^{\prime}\cdot\epsilon)U(\omega)+(\epsilon^{\prime}\cdot k)(\epsilon\cdot k)V(\omega).
\end{equation}
\begin{table}[htbp]
    \centering
    \captionsetup{justification=raggedright, singlelinecheck=false}
    \caption{Explicit expressions for the spin-averaged Compton scattering tensor $\Theta_{\mu\nu}^s$ for hadrons with spin $s$. Here, $m$ denotes the mass of the heavy hadron, and $P_{\alpha\beta}^{3/2} = g_{\alpha\beta} - v_{\alpha}v_{\beta} + \frac{4}{d-1}S_{\alpha}S_{\beta}$ represents the spin-3/2 projection operator.}
    \label{tab:Theta_tensor}
    \setlength{\tabcolsep}{2.5mm}
    \begin{tabular}{c|cccc}
    \hline\hline
        $s$ & $0$ & $\frac{1}{2}$ & $1$ & $\frac{3}{2}$   \\ \hline
        $\Theta^s_{\mu\nu}$ & $\frac{e^2}{2m} T_{\mu\nu}(v,k)$ & $\frac{e^2}{4}\mathrm{Tr}[(1+\slashed v) T_{\mu\nu}(v,k)]$ & $-\frac{e^2}{6m}(g_{\alpha\beta}-v_\alpha v_\beta)T^{\alpha\beta}_{\mu\nu}(v,k)$ & $-\frac{e^2}{8}\mathrm{Tr}[(1+\slashed v) P^{3/2}_{\alpha\beta} T^{\alpha\beta}_{\mu\nu}(v,k)]$  \\
        \hline\hline
    \end{tabular}
\end{table}

The general low-energy expansion of the Compton scattering amplitude in the laboratory frame (where $\vec{p}=0$) is given by:
\begin{equation}
	\mathcal{M}=\hat{\epsilon}\cdot\hat{\epsilon}^{\prime}\left(-\frac{Z^2 e^2}{M}+\omega\omega^{\prime}4\pi\alpha_E\right)+(\hat{\epsilon}\times\vec{k})\cdot(\hat{\epsilon}^{\prime}\times\vec{k}^{\prime})4\pi\beta_M+O\left(\omega^4\right),
\end{equation}
where $Z e$ is the hadron charge. By matching this expansion with Eq.~\eqref{eq:lorentzexp} and taking the forward case ($\vec{k} = \vec{k}'$, $\omega = \omega'$), the electric and magnetic polarizabilities are related to the structure functions $U(\omega)$ and $V(\omega)$ as~\cite{Llanta:1979kj,Bernabeu:1976jq,Bernard:1993bg,Bernard:1993ry}:
\begin{equation}\label{eq:polar}
\alpha_E+\beta_M=-\left.\frac{1}{8\pi}\frac{\partial^2}{\partial\omega^2}U(\omega)\right|_{\omega=0},\quad\beta_M=-\frac{1}{4\pi}V(\omega=0).\end{equation}
Our primary goal is to calculate all contributions to $U(\omega)$ and $V(\omega)$ up to \(\mathcal{O}(p^3)\) within the framework of HH$\chi$PT.
\subsection{Effective Chiral Lagrangian}\label{subsec:effective_Lagrangian}
We construct the effective Lagrangian for Goldstone bosons, heavy mesons with a heavy antiquark, and heavy baryons with two heavy quarks, all of which are the fundamental representations of $SU(3)_V$. For the non-linear realization of the chiral symmetry, we have
\begin{equation}
    U=u^2=e^{i \phi/F_\phi}
\end{equation}
where $\phi$ is the matrix for the Goldstone octet
\begin{equation}\phi=\sum_{a=1}^{8}\lambda_{a}\phi_{a}=
\begin{pmatrix}\label{pimeson}
\pi^{0}+\frac{1}{\sqrt{3}}\eta & \sqrt{2}\pi^{+} & \sqrt{2}K^{+} \\
\sqrt{2}\pi^{-} & -\pi^{0}+\frac{1}{\sqrt{3}}\eta & \sqrt{2}K^{0} \\
\sqrt{2}K^{-} & \sqrt{2}\bar{K}^{0} & -\frac{2}{\sqrt{3}}\eta
\end{pmatrix}.\end{equation}
We have ignored the mixing angle of $\pi_0$ and $\eta$ caused by the difference between the mass eigenstate and the flavor eigenstate. $F_\phi$ is the decay constant of the pseudoscalar meson in the chiral limit, where we will adopt $F_\pi=92.4$ MeV, $F_K=113$ MeV and $F_\eta=116$ MeV in this work.
We denote the pseudoscalar \((J^P=0^-)\) and vector \((J^P=1^-)\) ground states of a heavy meson composed of a heavy antiquark \(\bar{Q}\) and a light quark \(q\) by \(P\) and \(P^*_\mu\), respectively. For charmed mesons
\begin{equation}\label{eq:charm_meson}
P=(\bar{D}^0,D^-,D_s^-),\quad P_\mu^*=(\bar{D}_\mu^{*0},D_\mu^{*-},D_{s\mu}^{*-}).             
\end{equation}
and for bottom mesons
\begin{equation}\label{eq:bottom_meson}
P=(B^+,B^0,B_s^0),\quad P_\mu^*=(B_\mu^{*+},B_\mu^{*0},B_{s\mu}^{*0}).
\end{equation}
Similarly, a doubly charmed baryon composed of two heavy quarks and a light quark can be expressed as
\begin{align}
    \mathcal{B}=(\Xi_{cc}^{++},\Xi_{cc}^{+},\Omega_{cc}^{+}),\quad \mathcal{B}_\mu^*=(\Xi_{cc\mu}^{*++},\Xi_{cc\mu}^{*+},\Omega_{cc\mu}^{*+}),
\end{align}
where $\mathcal{B}$ and $\mathcal{B}^*$ are heavy baryons with spin-${1}/{2}$ and spin-${3}/{2}$, respectively. 
Under the $SU(3)_L\times SU(3)_R$ chiral transformation, these hadron fields transformed as
\begin{align}\label{trans}
    U&\to LUR^\dagger\nonumber\\
    \mathcal{H}&\to K\mathcal{H}\nonumber\\
    \mathcal{B}&\to K\mathcal{B}
\end{align}
where $R$ and $L$ are $SU(3)_R$ and $SU(3)_L$ transformation matrices, respectively. Gauge transformation $K$ is defined by the Goldstone fields in chiral transformation
\begin{equation}u\to RuK^\dagger=KuL^\dagger.\end{equation}
Therefore, we obtain $K(L,R,U)=\sqrt{RUL^\dagger}^{-1}R\sqrt{U}$.

According to the gauge transformation in Eq.\eqref{trans}, we can define the chiral connection and axial-vector current,
\begin{align}
\Gamma_{\mu}&\equiv\frac{1}{2}\left[u^\dagger\left(\partial_\mu-ir_\mu\right)u+u\left(\partial_\mu-il_\mu\right)u^\dagger\right],\nonumber\\
u_{\mu}&\equiv\frac{i}{2}\left[u^{\dagger}\left(\partial_{\mu}-ir_{\mu}\right)u-u\left(\partial_{\mu}-il_{\mu}\right)u^{\dagger}\right],\nonumber
\end{align}
where electromagnetic fields are introduced as the external fields of the left-hand and right-hand $r_\mu=l_\mu=-eQ_qA_\mu$, and $q=l,c(b),cc$ correspond to Goldstone, singly heavy mesons and doubly heavy baryons, respectively. Thus, the covariant derivatives of these hadron fields are defined as follows:
\begin{alignat}{3}
&\nabla_{\mu}U=\partial_{\mu}U-ir_{\mu}U+iUl_{\mu}\quad &&Q_l=\mathrm{diag}(2/3,-1/3,-1/3),\nonumber\\
&D_{\mu}\mathcal{H}=\partial_{\mu}\mathcal{H}+\Gamma_{\mu}\mathcal{H} \quad &&Q_c=\mathrm{diag}(0,-1,-1) \quad Q_b=\mathrm{diag}(1,0,0),\nonumber\\
&D_{\mu}\mathcal{B}=\partial_{\mu}\mathcal{B}+\Gamma_{\mu}\mathcal{B} \quad &&Q_{cc}=\mathrm{diag}(2,1,1).\nonumber
\end{alignat}
The chiral covariant electromagnetic field strength tensors $F^\pm_{\mu\nu}$ are defined as
\begin{align}
 & F_{\mu\nu}^{\pm}=u^\dagger F_{\mu\nu}^Ru\pm uF_{\mu\nu}^Lu^\dagger, \nonumber\\
 & F_{\mu\nu}^R=\partial_\mu r_\nu-\partial_\nu r_\mu-i\left[r_\mu,r_\nu\right], \nonumber\\
 & F_{\mu\nu}^L=\partial_\mu l_\nu-\partial_\nu l_\mu-i\left[l_\mu,l_\nu\right].\nonumber
\end{align}

Then we can construct the effective chiral Lagrangian for these hadron fields. The leading order (LO) Lagrangian for the Goldstone boson field is
\begin{equation}\mathcal{L}_{\phi\phi}^{(2)}=\frac{F_\phi^2}{4}\mathrm{Tr}[\nabla_\mu U\left(\nabla^\mu U\right)^\dagger]+\frac{F_\phi^2}{4}\mathrm{Tr}[\chi U+U\chi^\dagger],\end{equation}
where $\chi=2B_0\operatorname{diag}\left(m_u,m_d,m_s\right)$ denotes the chiral symmetry breaking effect of the current quark mass $m_{u,d,s}$ and $B_0$ is a parameter related to the quark condensate. We use $\mathrm{Tr}(X)$ and $\left\langle X\right\rangle $ to denote the trace of $X$ in the flavor space and the spinor space, respectively. 

In the HH$\chi$PT formalism, the LO Lagrangian for describing the interactions of heavy mesons reads:
\begin{equation}
    \mathcal{L}_{H\phi}^{(1)}=2iP^\dagger v\cdot D P-2P^{*\dagger}(iv\cdot D-\Delta)P^{*}+2g(iP^{*\dagger}_\mu u^\mu P+\mathrm{H.c.})-2\tilde{g}i\epsilon^{\mu \nu \rho \sigma}u_\mu P^{*\dagger}_\nu P^{*}_\rho v_\sigma.
\end{equation}
where $g$ and $\tilde{g}$ are the axial coupling constants and $\Delta=m_{P^*}-m_P$ stands for mass splitting. It is easy to observe that the vertex $\phi P P$ is forbidden due to angular momentum conservation. Under the constraints of heavy quark symmetry, the quark model yields $g = \tilde{g}$~\cite{PhysRevD.46.1148}. For the sake of brevity, we adopt the notation $g$ for both constants in the following discussion. The next-to-leading order (NLO) Lagrangian is
\begin{align}\mathcal{L}_{H\gamma}^{(2)}=&-P^{\dagger}\frac{D^{2}}{m_P}P+P^{*\dagger}\frac{D^{2}}{m_{P^*}}P^{*}+4i\tilde{a}P^{*\dagger\mu}P^{*\nu}\tilde{F}_{\mu\nu}^++4iaP^{*\dagger\mu}P^{*\nu}\operatorname{Tr}(F_{\mu\nu}^+)\nonumber\\
+&2i\tilde{a}\epsilon^{\mu\nu\rho\sigma} P^{*\dagger}_{\rho}Pv_\sigma\tilde{F}_{\mu\nu}^++\mathrm{H.c.}+2ia\epsilon^{\mu\nu\rho\sigma} P^{*\dagger}_{\rho}Pv_\sigma\operatorname{Tr}(F_{\mu\nu}^+)+\mathrm{H.c.},\end{align}
where $\tilde{a}$ and $a$ correspond to the LECs for the light quark and the heavy antiquark, respectively. $\tilde{F}_{\mu\nu}^\pm=F_{\mu\nu}^\pm-\frac{1}{3}\mathrm{Tr}(F_{\mu\nu}^\pm)$ related to the charge matrix of the light quarks is traceless. 

For baryons containing two identical heavy quarks ($ccq$ and $bbq$), the ground states are dominated by an $S$-wave configuration of the heavy-quark pair. The Pauli exclusion principle then dictates that the two heavy quarks in a color $\bar{\mathbf{3}}$ state must form a diquark with spin $S=1$.  Coupled with the light quark, this yields a ground-state doublet consisting of a spin-$1/2$ baryon $\mathcal{B}$ and a spin-$3/2$ baryon $\mathcal{B}^*_\mu$. The LO Lagrangian is written as~\cite{Meng:2017dni,PhysRevD.96.076011,Li:2017pxa}
\begin{align}
\mathcal{L}_{\mathcal{B}\phi}^{(1)}= & \bar{\mathcal{B}}iv\cdot D\mathcal{B}-\bar{\mathcal{B}}^*\left(iv\cdot D-\delta \right)\mathcal{B}^*
  +2g_1\bar{\mathcal{B}}S\cdot u\mathcal{B}+g_2(\bar{\mathcal{B}}_{\mu}^*u^\mu\mathcal{B}+\mathrm{~H.c.~})+2g_3\bar{\mathcal{B}}^*S\cdot u\mathcal{B}^*,
\end{align}
where $\delta$ represents the average mass differences between $\mathcal{B}^*$ and $\mathcal{B}$ and the parameters $g_i$ denote the coupling constants. The NLO Lagrangian of baryons is
\begin{align}
    \mathcal{L}_{\mathcal{B}\phi}^{(2)}&=\bar{\mathcal{B}}\frac{\left(v\cdot D\right)^{2}-D^{2}}{2M}\mathcal{B}-\bar{\mathcal{B}}^{*\mu}\frac{\left(v\cdot D\right)^{2}-D^{2}}{2M^{*}}\mathcal{B}_{\mu}^{*}-\frac{i\tilde{f}_1}{4M_N}\bar{\mathcal{B}}\left[S^\mu,S^\nu\right]\tilde{F}_{\mu\nu}^+\mathcal{B}-\frac{if_1}{4M_N}\bar{\mathcal{B}}\left[S^\mu,S^\nu\right]\mathcal{B} \mathrm{Tr}( F_{\mu\nu}^+)\nonumber\\
    &+\frac{i\tilde{f}_2}{4M_{N}}\bar{\mathcal{B}}\tilde{F}_{\mu\nu}^{+}S^{\nu}\mathcal{B}^{*\mu}+\mathrm{~H.c.~}+\frac{if_2}{4M_{N}}\bar{\mathcal{B}}S^{\nu}\mathcal{B}^{*\mu}\mathrm{Tr}(F_{\mu\nu}^{+})+\mathrm{H.c.}+\frac{i\tilde{f}_3}{4M_N}\bar{\mathcal{B}}^{*\mu}\tilde{F}_{\mu\nu}^+\mathcal{B}^{*\nu}+\frac{if_3}{4M_N}\bar{\mathcal{B}}^{*\mu}\mathcal{B}^{*\nu}\mathrm{Tr}(F_{\mu\nu}^+),
\end{align}
where $f_i$ and $\tilde{f}_i$ are LECs. We use the nucleon mass $M_N$ to render them dimensionless so that they can be expressed in units of the nuclear magneton. 

For the heavy baryons composed of three non-identical quarks ($bcq$), the internal structure involves distinct hierarchies in coordinate and spin spaces. While the two heavy quarks ($b$ and $c$) tend to form a compact spatial core due to the heavy quark interaction, the spin dynamics are governed by the color-magnetic hyperfine splitting, which scales inversely with the constituent masses. Consequently, the spin interaction between the charm and light quarks is significantly stronger than that between the heavy quarks. Recent quark model calculations~\cite{Ma:2022vqf} support this hyperfine hierarchy, indicating that the mass eigenstates are best described by the $cq$ spin-clustering scheme—where the charm and light quarks are first coupled in the spin basis—rather than the $bc$ spin-diquark structure. Based on this scheme, the two low-lying spin-$1/2$ states are identified by the spin of the $cq$ pair. The lower-energy state, denoted as $\mathcal{T}$, corresponds to the scalar $cq$ configuration ($S_{[cq]}=0$). The higher-energy state, denoted as $\mathcal{B}$, corresponds to the axial-vector $cq$ configuration ($S_{\{cq\}}=1$). The latter naturally forms a doublet with the spin-$3/2$ baryon $\mathcal{B}^*_\mu$ under heavy quark spin symmetry. Consequently, the effective Lagrangian must describe the dynamics of both the singlet $\mathcal{T}$ and the doublet ($\mathcal{B}, \mathcal{B}^*$), including the mixing between them. The LO and NLO Lagrangians are given by:

\begin{align}
\mathcal{L}_{\mathcal{B}\phi}^{(1)}&=\bar{\mathcal{T}}iv\cdot D\mathcal{T}
  +\bar{\mathcal{B}}(iv\cdot D-\delta_1)\mathcal{B}-\bar{\mathcal{B}}^*\left(iv\cdot D-\delta_2 \right)\mathcal{B}^*
  +2g_1\bar{\mathcal{B}}S\cdot u\mathcal{B}+g_2(\bar{\mathcal{B}}_{\mu}^*u^\mu\mathcal{B}+\mathrm{~H.c.~})+2g_3\bar{\mathcal{B}}^*S\cdot u\mathcal{B}^*\nonumber\\
  &+2c_1\bar{\mathcal{T}}S\cdot u\mathcal{T}+2c_2(\bar{\mathcal{T}}S\cdot u\mathcal{B}+\mathrm{~H.c.~})+c_3(\bar{\mathcal{B}}_{\mu}^*u^\mu\mathcal{T}+\mathrm{~H.c.~}),
\end{align}
\begin{align}\label{eq:bcq_lagrangian_NLO}
  \mathcal{L}_{\mathcal{B}\phi}^{(2)}&=\bar{\mathcal{T}}\frac{\left(v\cdot D\right)^{2}-D^{2}}{2M_{\mathcal{T}}}\mathcal{T}+\bar{\mathcal{B}}\frac{\left(v\cdot D\right)^{2}-D^{2}}{2M}\mathcal{B}-\bar{\mathcal{B}}^{*\mu}\frac{\left(v\cdot D\right)^{2}-D^{2}}{2M^{*}}\mathcal{B}_{\mu}^{*}-\frac{i\tilde{f}_1}{4M_N}\bar{\mathcal{B}}\left[S^\mu,S^\nu\right]\tilde{F}_{\mu\nu}^+\mathcal{B}\nonumber\\
  &-\frac{if_1}{4M_N}\bar{\mathcal{B}}\left[S^\mu,S^\nu\right]\mathcal{B} \mathrm{Tr}( F_{\mu\nu}^+)\nonumber+\frac{i\tilde{f}_2}{4M_{N}}\bar{\mathcal{B}}\tilde{F}_{\mu\nu}^{+}S^{\nu}\mathcal{B}^{*\mu}+\mathrm{~H.c.~}+\frac{if_2}{4M_{N}}\bar{\mathcal{B}}S^{\nu}\mathcal{B}^{*\mu}\mathrm{Tr}(F_{\mu\nu}^{+})+\mathrm{H.c.}\nonumber\\
    &+\frac{i\tilde{f}_3}{4M_N}\bar{\mathcal{B}}^{*\mu}\tilde{F}_{\mu\nu}^+\mathcal{B}^{*\nu}+\frac{if_3}{4M_N}\bar{\mathcal{B}}^{*\mu}\mathcal{B}^{*\nu}\mathrm{Tr}(F_{\mu\nu}^+),-\frac{i\tilde{f}_4}{4M_N}\bar{\mathcal{T}}\left[S^\mu,S^\nu\right]\tilde{F}_{\mu\nu}^+\mathcal{T}-\frac{if_4}{4M_N}\bar{\mathcal{T}}\left[S^\mu,S^\nu\right]\mathcal{T} \mathrm{Tr}( F_{\mu\nu}^+)\nonumber\\
    &-\frac{i\tilde{f}_5}{4M_N}\bar{\mathcal{B}}\left[S^\mu,S^\nu\right]\tilde{F}_{\mu\nu}^+\mathcal{T}-\mathrm{H.c.}-\frac{if_5}{4M_N}\bar{\mathcal{B}}\left[S^\mu,S^\nu\right]\mathcal{T} \mathrm{Tr}( F_{\mu\nu}^+)-\mathrm{H.c.}+\frac{i\tilde{f}_6}{4M_{N}}\bar{\mathcal{T}}\tilde{F}_{\mu\nu}^{+}S^{\nu}\mathcal{B}^{*\mu}+\mathrm{~H.c.~}\nonumber\\
    &+\frac{if_6}{4M_{N}}\bar{\mathcal{T}}S^{\nu}\mathcal{B}^{*\mu}\mathrm{Tr}(F_{\mu\nu}^{+})+\mathrm{H.c.}
\end{align}
where $\delta_1$ and $\delta_2$ denote the mass differences of $\mathcal{B}$ and $\mathcal{B}^*$ relative to the singlet state $\mathcal{T}$. The coupling constants $c_i$ characterize the interactions involving the singlet field, while $g_i$ describe the interactions within the doublet.

\begin{figure}
    \centering
    \includegraphics[width=1\linewidth]{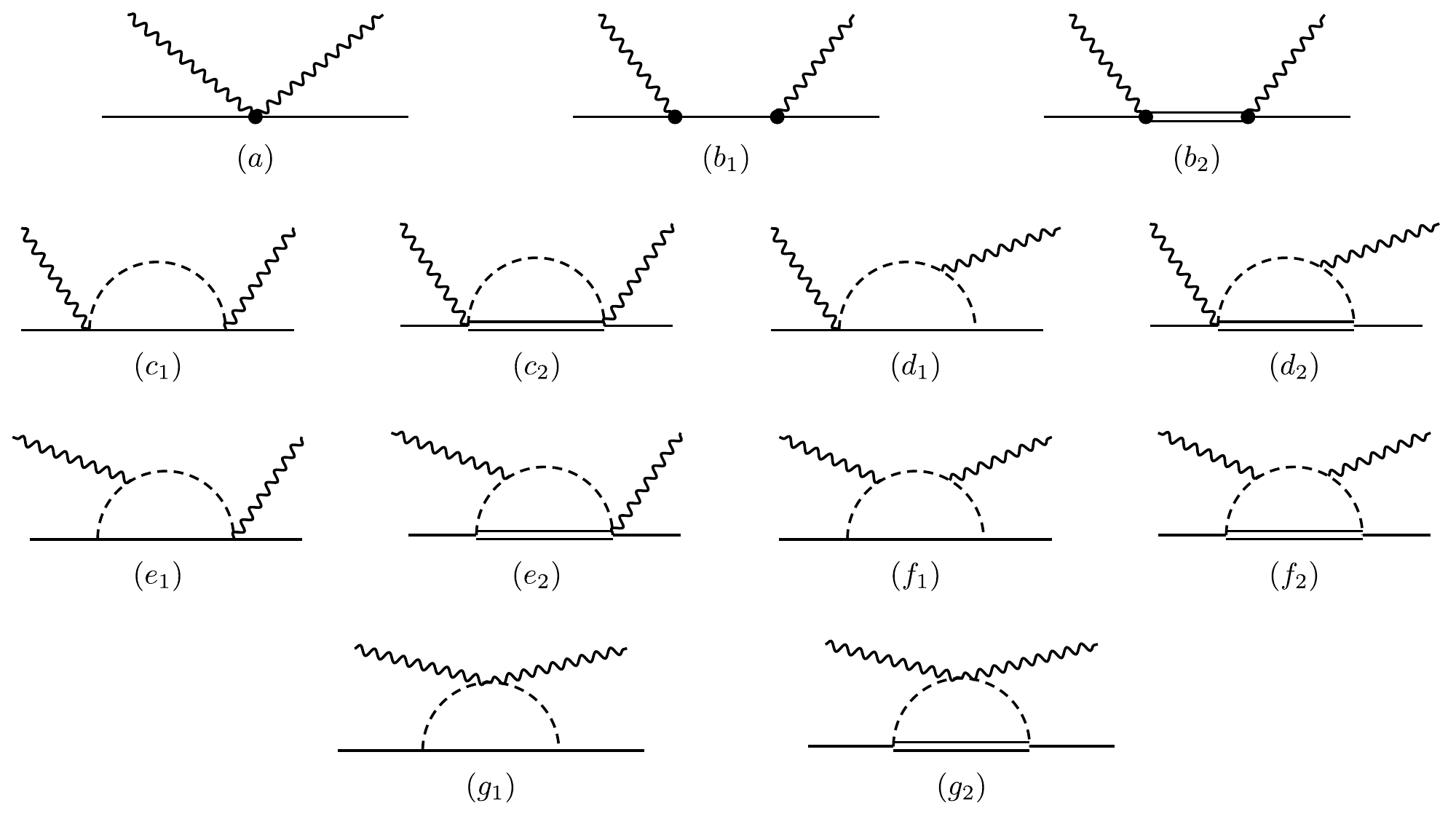}
    \captionsetup{justification=raggedright, singlelinecheck=false}
    \caption{The Born and loop diagrams contribute to the electromagnetic polarizabilities of pseudoscalar mesons (spin-${1}/{2}$ baryons) up to \(\mathcal{O}(p^3)\). The solid dots denote the $\mathcal{L}_{H\gamma}^{(2)}$ ($\mathcal{L}_{B\phi}^{(2)}$) vertices. The single and double lines represent the pseudoscalar mesons (spin-${1}/{2}$ baryons) and vector mesons (spin-${3}/{2}$ baryons), respectively. Crossed diagrams are not shown.}
    \label{fig:spin0}
\end{figure}

\begin{figure}
    \centering
    \includegraphics[width=1\linewidth]{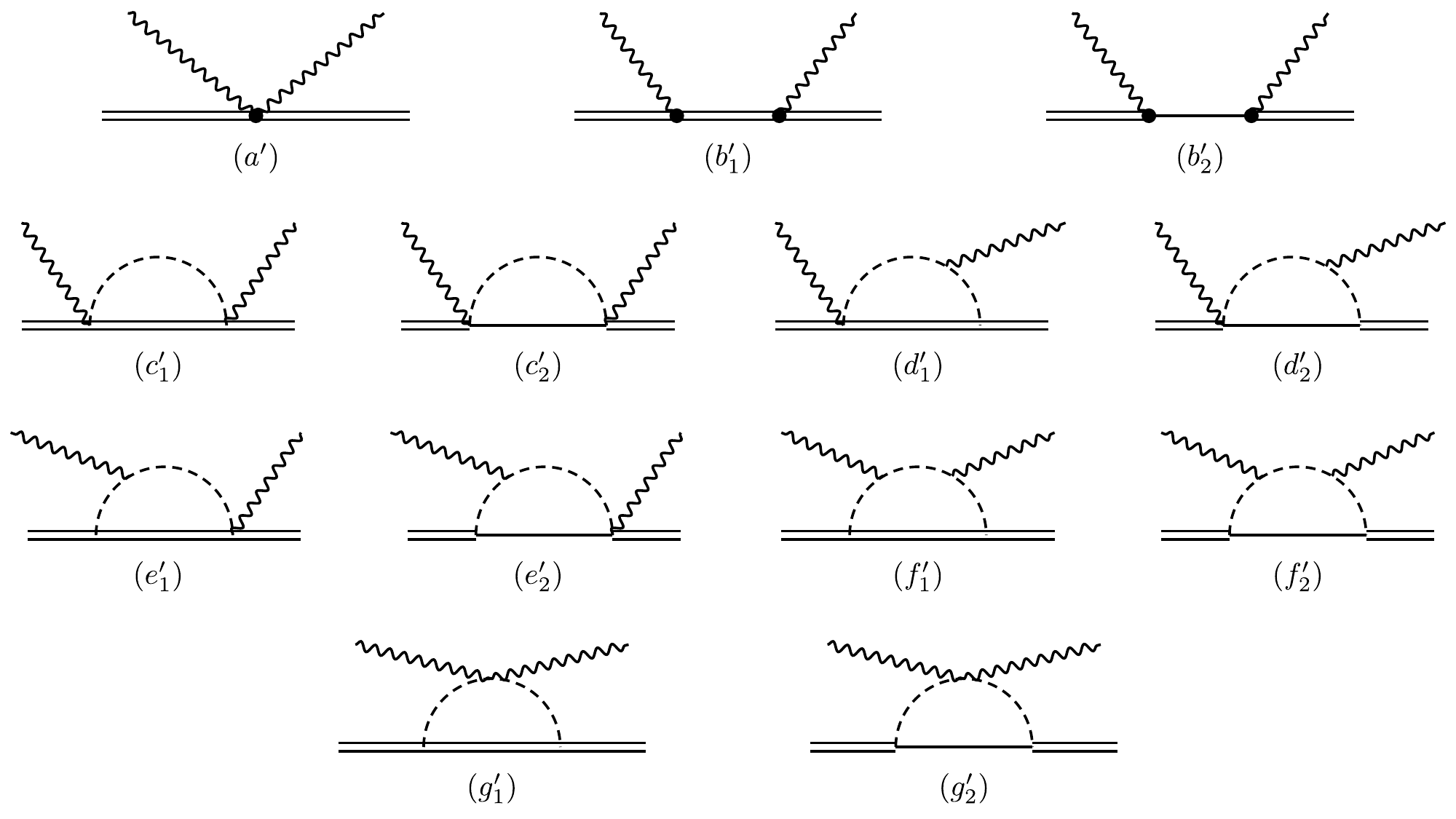}
    \captionsetup{justification=raggedright, singlelinecheck=false}
    \caption{The Born and loop diagrams contribute to the electromagnetic polarizabilities of vector mesons (spin-${3}/{2}$ baryons) up to \(\mathcal{O}(p^3)\). The notations are the same as those in Fig.~\ref{fig:spin0}.}
    \label{fig:spin1}
\end{figure}

Due to the crossing symmetry, the forward Compton scattering is even in the photon momentum, the \(\mathcal{O}(p^3)\) of Lagrangians does not contribute to the polarizabilities, which is omitted in this work. Besides, the “Coulomb gauge” makes the $\gamma\gamma P(\mathcal{B})$ vertex proportional to $\epsilon\cdot v$ vanish, significantly reducing the number of diagrams that need to be calculated. The tree and loop Feynman diagrams that contribute to the electromagnetic polarizabilities of pseudoscalar mesons (spin-${1}/{2}$ baryons) and vector mesons (spin-${3}/{2}$ baryons) up to \(\mathcal{O}(p^3)\) are shown in Fig.~\ref{fig:spin0} and Fig.~\ref{fig:spin1}, respectively.

\section{Electromagnetic Polarizabilities of Singly Heavy Mesons}\label{sec:meson}
In this section, we present the calculation of the electromagnetic polarizabilities for singly heavy mesons up to $\mathcal{O}(p^3)$ within the HH$\chi$PT framework. We first analyze the tree-level contributions. The $\mathcal{O}(p^2)$ seagull diagrams, shown in Fig.~\ref{fig:spin0}($a$) and Fig.~\ref{fig:spin1}($a^\prime$), generate the standard Thomson scattering amplitudes
\begin{alignat}{2}\label{eq:thomson1}
U_P^{(a)}(\omega)&=\frac{Q_P^2e^2}{m_P},\quad U_{P^*}^{(a^\prime)}(\omega)=\frac{Q_{P^*}^2e^2}{m_{P^*}},\nonumber\\
 V_P^{(a)}(\omega)&=V_{P^*}^{(a^\prime)}(\omega)=0,
\end{alignat}
These terms describe the scattering of a charged point particle and do not contribute to the polarizabilities, which characterize particle deformation. Consequently, we have $\alpha_E^{(a)}(P) = \beta_M^{(a)}(P)=\alpha_E^{(a')}(P^*) = \beta_M^{(a')}(P^*) = 0$.

The \(\mathcal{O}(p^3)\) tree diagram in Fig.~\ref{fig:spin0}($b_1$) and Fig.~\ref{fig:spin1}($b^\prime_1$) cancels exactly with its crossed counterpart. As a result, these diagrams do not contribute to the electromagnetic polarizabilities. 
However, the magnetic transition diagrams involving the $P P^* \gamma$ vertex, shown in Fig.~\ref{fig:spin0}$(b_2)$ and Fig.~\ref{fig:spin1}$(b^{\prime}_2)$, provide finite contributions. These diagrams correspond to two consecutive magnetic dipole transitions and contribute exclusively to the magnetic polarizability $\beta_M$. The non-vanishing structure functions are
\begin{equation}
U_P^{(b_2)}(\omega)=-32e^2 C_P^2\omega^2 \frac{\Delta}{\Delta^2-\omega^2},\quad V_P^{(b_2)}(\omega)=-32e^2 C_P^2\frac{\Delta}{\Delta^2-\omega^2},
\end{equation}
\begin{equation}
U_{P^*}^{(b^{\prime}_2)}(\omega)=\frac{32e^2 C_{P^*}^2\omega^2}{3 }\frac{\Delta}{\Delta^2-\omega^2},\quad V_{P^*}^{(b^{\prime}_2)}(\omega)=\frac{32e^2 C_{P^*}^2}{3 }\frac{\Delta}{\Delta^2-\omega^2},
\end{equation}
where $C_P$ and $C_{P^*}$ collect the effective coupling constants and flavor factors derived from the Lagrangian $\mathcal{L}_{H\gamma}^{(2)}$. In the heavy quark limit, spin-flavor symmetry dictates $C_{D}=C_{B}=C_{D^{*}}=C_{B^{*}}$. Explicitly, for the charmed sector, we have $$C_{\bar{D}^0}=\frac{2}{3}\tilde{a}-2a,\quad C_{D^-}=-\frac{1}{3}\tilde{a}-2a,\quad C_{D_s^-}=-\frac{1}{3}\tilde{a}-2a.$$ Using Eq.\eqref{eq:polar}, we extract the tree-level magnetic polarizabilities as
\begin{align}
\beta_{M}^{(b_{2})}(P)=\frac{8\alpha_{em}C_{P}^{2}}{M_{N}^{2}}\frac{1}{\Delta}, \quad \beta_{M}^{(b_{2}^{\prime})}(P^{*})=-\frac{8\alpha_{em}C_{P^*}^{2}}{3M_{N}^{2}}\frac{1}{\Delta}, \quad
\end{align}
while the electric polarizabilities vanish at this order, $\alpha_{E}^{(b_{2})}(P)=\alpha_{E}^{(b_{2}^{\prime})}(P^{*})=0$.

The dominant contributions to the electric polarizability arise from the chiral loop diagrams shown in Figs.~\ref{fig:spin0}($c$)--($g$) and Figs.~\ref{fig:spin1}($c'$)--($g'$). These loops capture the non-perturbative physics of the pion cloud surrounding the heavy hadron. For the pseudoscalar mesons, the absence of a $PP\phi$ vertex implies that the electric polarizability is generated entirely by the $P^*\phi$ loops. For the vector mesons, both $P\phi$ and $P^*\phi$ loops contribute. The analytical expressions for the loop functions are provided in Appendix~\ref{appendix:expression_of_EMP}.  It is instructive to note that in the heavy quark limit ($\Delta \to 0$), the loop contributions satisfy the relation 
\begin{equation}
    \alpha_E^{(c^\prime_1-g^\prime_1)}(P^*)=2\alpha_E^{(c^\prime_2-g^\prime_2)}(P^*), \quad \beta_M^{(c^\prime_1-g^\prime_1)}(P^*)=2\beta_M^{(c^\prime_2-g^\prime_2)}(P^*),
\end{equation}
reflecting the underlying heavy quark spin symmetry. 

\subsection{NUMERICAL RESULTS}

Up to \(\mathcal{O}(p^3)\), three LECs enter the calculation: the axial coupling constants $g$ from $\mathcal{L}_{H\phi}^{(1)}$ and the magnetic dipole transition parameters $a(\tilde{a})$ from $\mathcal{L}_{H\gamma}^{(2)}$. For $D$ mesons, the axial coupling $g$ can be extracted by the partial decay width of $D^{*+}$~\cite{CLEO:2001foe},
\begin{equation}
\Gamma(D^{*+}\to D^0 \pi^+)=\frac{g^2}{12\pi F_{\pi}^2}|\vec{p}_{\pi^+}|^3.
\end{equation}
For $B$ mesons, the corresponding axial coupling $g_b$ is taken from lattice QCD determinations~\cite{Ohki:2008py}. In this work, we adopt
\begin{equation}
    g_c=0.59\pm0.01\pm0.07,\quad g_b=0.516\pm0.05\pm0.033.\nonumber
\end{equation}
where the subscripts $c$ and $b$ refer to the charmed or bottom mesons, respectively. The magnetic dipole transition parameters $a$ and $\tilde{a}$ are estimated using quark model calculations~\cite{Wang:2019mhm},
\begin{equation}
    \tilde{a}=-\frac{1}{8m_q},\quad a=\frac{1}{24m_{\bar{Q}}},
\end{equation}
where $m_q$ and $m_{\bar{Q}}$ are the masses of the light quark and heavy antiquark that constitute a specific meson, respectively. We adopt the same values as in previous HH\(\chi\)PT studies~\cite{Wang:2019mhm,Li:2017pxa,PhysRevD.96.076011,Meng:2017dni,PhysRevD.98.054026}, as summarized in Table~\ref{tab:quark_mass}, and conservatively assign a $10\%$ uncertainty to $a(\tilde{a})$. Notably, since the mass splitting between \(D\) and \(D^*\) is comparable to the pion mass, the electric polarizabilities of \(D^*\) are highly sensitive to these values. We therefore evaluate the mass difference using the physical masses quoted by the PDG (averaged values)~\cite{ParticleDataGroup:2024cfk}. 

\begin{table}[htbp]
    \centering
    \caption{The masses of the constituent quarks (in units of $\text{GeV}$).}
    \label{tab:quark_mass}
    \setlength{\tabcolsep}{2.5mm}
    \begin{tabular}{ccccc}
    \hline\hline
         $m_u$& $m_d$ & $m_s$ & $m_c$ & $m_b$  \\ \hline
        $0.336$ & $0.336$ & $0.540$ & $1.660$ & $4.730$   \\
        \hline\hline
    \end{tabular}
\end{table}

The numerical results for the electromagnetic polarizabilities of singly heavy mesons are shown in Table~\ref{tab:polarizabilities_numerical_results_meson}. The uncertainties are derived through standard error propagation. They originate from the experimental uncertainties in the axial coupling constants $g$ and the physical meson masses, as well as the theoretical uncertainty associated with the constituent quark masses. For pseudoscalar mesons, we observe that $\alpha_E^{(c_1-g_1)}(P)=\beta_M^{(c_1-g_1)}(P)=0$ due to the absence of $\phi P P$ vertex. The results indicate that the long‑range chiral corrections from the \(P^*\phi\)-loop shown in Fig.~\ref{fig:spin0}($c_2$)-($g_2$) provide the dominant contribution to the electric polarizabilities, whereas the magnetic-dipole transition $P^*\to P \gamma$ shown in Fig.~\ref{fig:spin0}($b_2$) mainly determines the magnetic polarizabilities.

\begin{table}[htbp]
    \captionsetup{justification=raggedright, singlelinecheck=false}
    \caption{The numerical results of singly heavy meson electromagnetic polarizabilities (in unit of $10^{-4}~\mathrm{fm}^3$). The values in parentheses represent the uncertainties of the results.}
	\label{tab:polarizabilities_numerical_results_meson}
    \centering
    \setlength{\tabcolsep}{1.5mm}
    \begin{tabular}{c|ccc|cccc}
    \hline\hline
         & $\alpha_E^{(c_1-g_1)}$ & $\alpha_E^{(c_2-g_2)}$ & $\alpha_E^{\text{Tot.}}$ & $\beta_M^{(b_2)}$ & $\beta_M^{(c_1-g_1)}$ & $\beta_M^{(c_2-g_2)}$ &  $\beta_M^{\mathrm{Tot.}}$   \\ \hline
        $\bar{D}^{0}$ & $0$ & $1.50(36)$ & $1.50(36)$ & $11.24(191)$ & $0$ & $0.21(5)$ & $11.45(191)$ \\ 
        $D^{-}$ & $0$ & $1.22(29)$ & $1.22(29)$ & $0.69(25)$ & $0$ & $0.17(4)$ & $0.87(26)$ \\ 
        $D_s^{-}$ & $0$ & $0.47(11)$ & $0.47(11)$ & $0.09(6)$ & $0$ & $0.05(1)$ & $0.14(7)$ \\
        $B^{+}$ & $0$ & $1.78(41)$ & $1.78(41)$ & $22.64(470)$ & $0$ & $0.21(5)$ & $22.85(470)$ \\ 
        $B^{0}$ & $0$ & $1.49(35)$ & $1.49(35)$ & $7.01(131)$ & $0$ & $0.17(4)$ & $7.19(131)$ \\ 
        $B_s^{0}$ & $0$ & $0.44(10)$ & $0.44(10)$ & $2.71(50)$ & $0$ & $0.04(1)$ & $2.75(50)$ \\
        \hline
        & $\alpha_E^{(c'_1-g'_1)}$ & $\alpha_E^{(c'_2-g'_2)}$ & $\alpha_E^{\text{Tot.}}$ & $\beta_M^{(b'_2)}$ & $\beta_M^{(c'_1-g'_1)}$ & $\beta_M^{(c'_2-g'_2)}$ &  $\beta_M^{\mathrm{Tot.}}$   \\ \hline
        $\bar{D}^{*0}$ & $2.15(52)$ & $291(105)$ & $294(106)$ & $-3.75(64)$ & $0.22(5)$ & $0.98(25) $ & $-2.55(70)$\\ 
        $D^{*-}$ & $1.81(43)$ & \makecell[c]{-0.39(9) \\ -64.5(155)i} & \makecell[c]{1.42(34) \\ -64.5(155)i} & $-0.23(8)$ & $0.18(4)$ & \makecell[c]{-0.06(1) \\ +0.62(15)i} & \makecell[c]{-0.11(4) \\ +0.62(12)i} \\ 
        $D_s^{*-}$ & $0.34(8)$ & $0.39(9)$ & $0.74(18)$ & $-0.03(2)$ & $0.03(1)$ & $0.03(1)$ & $0.03(3)$ \\  
        $B^{*+}$ & $1.65(38)$ & $1.24(29)$ & $2.89(67)$ & $-7.55(157)$ & $0.16(4)$ & $0.10(2)$ & $-7.28(157)$\\ 
        $B^{*0}$ & $1.38(32)$ & $1.12(26) $ & $2.51(58) $ & $-2.34(44)$ & $0.14(3)$ & $0.09(2)$ & $-2.11(44)$ \\ 
        $B_s^{*0}$ & $0.26(6)$ & $0.19(5)$ & $0.46(11)$ & $-0.90(17)$ & $0.03(1)$ & $0.02$ & $-0.86(17)$ \\
        \hline\hline
    \end{tabular}
\end{table}

For vector mesons, the situation is different. A striking feature of our results is that the electric polarizabilities of $\bar{D}^{*0}$ and $D^{*-}$ are enhanced by several orders of magnitude compared to their bottom counterparts. This enhancement originates from the threshold behavior of chiral loop contributions. In the charmed sector, the mass splitting $\Delta \approx 142$ MeV is remarkably close to the pion mass $m_\pi \approx 140$ MeV. This near-degeneracy ($\Delta \sim m_\pi$) causes the energy denominators in the loop functions to vanish, generating a threshold singularity. 
Two related effects emerge from this. First, the loop amplitudes develop a nonvanishing imaginary part due to the opening of the $\bar{D}^{0}\pi^{-}$ channel, which contributes to the imaginary part of the polarizabilities of $D^{*-}$. Second, the virtual pion cloud surrounding the $D^{*}$ meson becomes weakly bound and spatially extended. This renders the system highly susceptible to external electromagnetic fields, resulting in strongly enhanced electric polarizability. The absence of a similar enhancement for the pseudoscalar $D$ meson, despite the loop integrals involving the same mass difference $\Delta \approx m_\pi$, stems from the distinct kinematic configuration. For the $D$ meson, the chiral loop is mediated by the heavier vector meson $D^*$. This intermediate state is significantly off-shell, leading to a suppression in the numerator of the loop amplitude that counteracts the small energy denominator. Consequently, no threshold enhancement occurs. The scenario changes fundamentally when we turn to the bottom sector. In this case, the mass splitting $\Delta \ll m_{\pi}$ places the system far below the pion production threshold, thereby suppressing the threshold effects observed in the charmed counterparts. Furthermore, the magnetic polarizabilities arising from intermediate spin-0 heavy mesons, $\beta_M^{(b'_2)}$, are found to be negative. Although this contribution remains dominant, the cancellation between tree and loop diagrams would reduce the total magnetic polarizabilities.

\section{Electromagnetic Polarizabilities of Doubly Heavy Baryons}\label{sec:baryon}
Similar to the case of mesons, the tree diagrams in Figs.~\ref{fig:spin0}$(a)$ and Figs.~\ref{fig:spin0}$(a^\prime)$ yield the Thomson amplitude:
\begin{alignat}{2}\label{eq:thomson}
U_\mathcal{B}^{(a)}(\omega)&=\frac{Q_\mathcal{B}^2e^2}{m_\mathcal{B}},\quad U_{\mathcal{B}^*}^{(a^\prime)}(\omega)=\frac{Q_{\mathcal{B}^*}^2e^2}{m_{\mathcal{B}^*}},\nonumber\\
 V_\mathcal{B}^{(a)}(\omega)&=V_{\mathcal{B}^*}^{(a^\prime)}(\omega)=0,
\end{alignat}
which does not contribute to the electromagnetic polarizabilities:
\begin{equation}
    \alpha_E^{(a)}(\mathcal{B})=\alpha_E^{(a^\prime)}({\mathcal{B}^*})=\beta_M^{(a)}(\mathcal{B})=\beta_M^{(a^\prime)}({\mathcal{B}^*})=0.
\end{equation}

For baryons containing two identical heavy quarks ($ccq$ and $bbq$), the $\mathcal{O}(p^{3})$ contributions from Fig.~\ref{fig:spin0}($b_{1}$) and Fig.~\ref{fig:spin1}($b_{1}^{\prime}$) vanish due to the exact cancellation with their crossed diagrams. However, the $bcq$ system, composed of non-identical heavy quarks, exhibits a distinct behavior. The mass splitting $\delta_1$ between the spin-1/2 ground states $\mathcal{T}$ and $\mathcal{B}$ induces a finite contribution through the intermediate transitions shown in Fig.~\ref{fig:spin0}($b_{1}$)
\begin{equation}
U_{\mathcal{T}}^{(b_1)}(\omega)=-\frac{e^2 C_{\mathcal{T}\mathcal{B}}^2\omega^2}{2 M_N^2}\frac{\delta_1}{\delta_1^2-\omega^2},\quad V_{\mathcal{T}}^{(b_1)}(\omega)=-\frac{e^2 C_{\mathcal{T}\mathcal{B}}^2}{2 M_N^2}\frac{\delta_1}{\delta_1^2-\omega^2},
\end{equation}
\begin{equation}
U_{\mathcal{B}}^{(b_1)}(\omega)=\frac{e^2 C_{\mathcal{B}\mathcal{T}}^2\omega^2}{2 M_N^2}\frac{\delta_1}{\delta_1^2-\omega^2},\quad V_{\mathcal{B}}^{(b_1)}(\omega)=\frac{e^2 C_{\mathcal{B}\mathcal{T}}^2}{2 M_N^2}\frac{\delta_1}{\delta_1^2-\omega^2}.
\end{equation}
Consequently, the resulting electromagnetic polarizabilities are:
\begin{align}
    \alpha_E^{(b_1)}({\mathcal{T}})&=\alpha_E^{(b_1)}({\mathcal{B}})=0,\nonumber\\
    \beta_M^{(b_1)}({\mathcal{T}})&=\frac{\alpha_{em} C_{\mathcal{T}\mathcal{B}}^2}{2 M_N^2}\frac{1}{\delta_1},\nonumber\\
    \beta_M^{(b_1)}({\mathcal{B}})&=-\frac{\alpha_{em} C_{\mathcal{B}\mathcal{T}}^2}{2 M_N^2}\frac{1}{\delta_1},
\end{align}

Here, $C_{\xi\xi^\prime}$ denotes the effective coupling strength for the magnetic dipole transition between baryon states $\xi$ and $\xi^\prime$. Due to time-reversal symmetry, these coefficients are symmetric, i.e., $C_{\xi\xi^\prime} = C_{\xi^\prime\xi}$. They are determined by the magnetic low-energy constants in the NLO Lagrangian Eq.~\eqref{eq:bcq_lagrangian_NLO} as follows:
$$C_{\mathcal{T}\mathcal{B}}=Q_q\tilde{f}_5+4f_5,\quad C_{\mathcal{T}\mathcal{B}^*} = Q_q\tilde{f}_6+4f_6,\quad  C_{\mathcal{B}\mathcal{B}^*}=Q_q\tilde{f}_2+4f_2,$$
where $Q_q$ represents the electric charge number of the light quark $q$ (specifically, $Q_u=2/3$ and $Q_d=Q_s=-1/3$). Using the same notation, the results of Fig.~\ref{fig:spin0}($b_2$) and Fig.~\ref{fig:spin1}($b'_2$) can be expressed in a unified form:
\begin{align}
    \alpha_E^{(b_2)}({\mathcal{T}})&=\alpha_E^{(b_2)}({\mathcal{B}})=\alpha_E^{(b^\prime_2)}({\mathcal{B}^*})=0\nonumber\\
    \beta_M^{(b_2)}({\mathcal{T}})&=\frac{\alpha_{em} C_{\mathcal{T}\mathcal{B}^*}^2}{12 M_N^2}\frac{1}{\delta_2}\nonumber\\
    \beta_M^{(b_2)}({\mathcal{B}})&=\frac{\alpha_{em} C_{\mathcal{B}\mathcal{B}^*}^2}{12 M_N^2}\frac{1}{\delta}\nonumber\\
    \beta_M^{(b^\prime_2)}({\mathcal{B}^*})&=-\frac{\alpha_{em} C_{\mathcal{B}^*\mathcal{T}}^2}{24 M_N^2}\frac{1}{\delta_2}-\frac{\alpha_{em} C_{\mathcal{B}^*\mathcal{B}}^2}{24 M_N^2}\frac{1}{\delta}
\end{align}
The results of loop diagrams are also shown in Appendix~\ref{appendix:expression_of_EMP}. 

\subsection{NUMERICAL RESULTS}
Since experimental data for doubly heavy baryons are currently unavailable, we employ the constituent quark model to estimate the relevant LECs~\cite{PhysRevD.46.1148}. We first address the axial coupling constants $g_i$ and $c_i$ appearing in the LO Lagrangian. For the $ccq$ and $bbq$ systems, the quark model relates their axial couplings directly to the nucleon axial charge $g_N \approx 1.267$~\cite{Meng:2017dni,PhysRevD.96.076011,Li:2017pxa},
\begin{equation}\label{eq:axial_coupling}
    g_1=-\frac{1}{5}g_N,\quad g_2=-\frac{2\sqrt{3}}{5}g_N,\quad g_3=-\frac{3}{5}g_N.
\end{equation}
For the $bcq$ system, the situation is similar but involves a more intricate internal structure. Following the quark model results~\cite{Ma:2022vqf}, the internal structure is characterized by the $cq$-clustering scheme, where the light quark is correlated with the charm quark. The ground states are identified as the scalar $b[cq]$ ($\mathcal{T}$) and the axial-vector $b\{cq\}$ ($\mathcal{B}, \mathcal{B}^*$). The axial couplings are estimated by evaluating the spin–flavor matrix elements within the quark model. The adopted values for the mass splittings and axial couplings are summarized in Table~\ref{tab:LEC_baryons}.

Subsequently, we estimate the LECs appearing in the NLO Lagrangian. It is crucial to note that the individual electromagnetic coupling constants $f_i$ and $\tilde{f}_i$ in the effective Lagrangian Eq.~\eqref{eq:bcq_lagrangian_NLO} are not determined separately. Instead, we directly fix the effective coupling strengths $C_{\xi\xi^\prime}$ by matching the magnetic dipole transitions derived from HH$\chi$PT to the predictions of the non-relativistic constituent quark model. This procedure allows us to express $C_{\xi\xi^\prime}$ in terms of the constituent quark magnetic moments $\mu_q$. For the baryons with two identical heavy quarks ($ccq$ and $bbq$), the effective couplings within the doublet ($\mathcal{B}, \mathcal{B}^*$) are given by:$$\begin{aligned}
C_{\mathcal{B}\mathcal{B}^{*}}(ccq) &= \frac{4\sqrt{3}}{3}(\mu_{q}-\mu_{c}), \\
C_{\mathcal{B}\mathcal{B}^{*}}(bbq) &= \frac{4\sqrt{3}}{3}(\mu_{q}-\mu_{b}).
\end{aligned}$$For the $bcq$ system, the mixing between the scalar ($S_{[cq]}=0$) and axial-vector ($S_{\{cq\}}=1$) configurations leads to three distinct transition coefficients:$$\begin{aligned}
C_{\mathcal{T}\mathcal{B}}(bcq) &= -\frac{1}{\sqrt{3}}(\mu_{q}-\mu_{c}), \\
C_{\mathcal{T}\mathcal{B}^{*}}(bcq) &= \frac{2\sqrt{3}}{3}(\mu_{c}+\mu_{q}-2\mu_{b}), \\
C_{\mathcal{B}\mathcal{B}^{*}}(bcq) &= 2(\mu_{q}-\mu_{c}).
\end{aligned}$$Here, $\mu_q$, $\mu_c$, and $\mu_b$ denote the magnetic moments of the light, charm, and bottom quarks, respectively. We adopt the numerical values from our previous works (in units of $\mu_N$)~\cite{Wen:2025xed,Chen:2024xks}:
\begin{equation}
	\mu_u=-2\mu_d=1.078(88),\quad \mu_s=-0.456(23),\quad \mu_c=0.205(15),\quad \mu_b=-0.05(5).
\end{equation}

\begin{table}[htbp]
	\centering
    \captionsetup{justification=raggedright, singlelinecheck=false}
	\caption{Numerical inputs for the axial coupling constants and mass splittings (in unit of MeV). The entries marked with ``-'' indicate couplings or splittings that are not defined for identical heavy quark systems.}
	\label{tab:LEC_baryons}
	\setlength{\tabcolsep}{2.5mm}
	\begin{tabular}{c|ccccccccc}
		\hline\hline
		& $g_1$ & $g_2$ & $g_3$ & $c_1$ & $c_2$ & $c_3$  & $\delta $ & $\delta_1$& $\delta_2$ \\ \hline
        $ccq$ & $-0.25(3)$ & $-0.88(9)$ & $-0.76(8)$ & $-$ & $-$& $-$ & $90$& $-$& $-$\\
		$bbq$ & $-0.25(3)$ & $-0.88(9)$ & $-0.76(8)$ & $-$ & $-$& $-$ & $37$& $-$& $-$\\
		$bcq$ & $0.51(5)$ & $-0.44(4)$ & $-0.76(8)$ & $0$ & $-0.44(4)$& $-0.76(8)$ & $31$& $41$& $72$ \\
		\hline\hline
	\end{tabular}
\end{table}

\begin{table}[htbp]	
    \captionsetup{justification=raggedright, singlelinecheck=false}
    \caption{The numerical results of doubly heavy baryon electromagnetic polarizabilities (in unit of $10^{-4}~\mathrm{fm}^3$). $b[cq]$ and $b\{cq\}$ denote scalar $S_{[cq]}=1$ state $\mathcal{T}$ and axial-vector $S_{\{cq\}}=1$ state $\mathcal{B}$, respectively. Uncertainties are given in parentheses.}
	\label{tab:polarizabilities_numerical_results_ccbaryon}
    \centering
    \setlength{\tabcolsep}{1.5mm}
    \begin{tabular}{c|ccc|ccccc}
    \hline\hline
         & $\alpha_E^{(c_1-g_1)}$ & $\alpha_E^{(c_2-g_2)}$ & $\alpha_E^{\text{Tot.}}$& $\beta_M^{(b_1)}$ & $\beta_M^{(b_2)}$ & $\beta_M^{(c_1-g_1)}$ & $\beta_M^{(c_2-g_2)}$ &  $\beta_M^{\mathrm{Tot.}}$   \\ \hline
        $\Xi_{cc}^{++}$ & $0.60(12)$ & $2.85(57)$ & $3.44(69)$& $0$ & $2.40(49)$ & $0.06(1)$ & $0.36(7)$ & $2.82(50)$ \\ 
        $\Xi_{cc}^{+}$ & $0.50(10)$ & $2.23(45)$ & $2.73(55)$& $0$ & $1.74(22)$ & $0.05(1)$ & $0.29(6)$ & $2.08(23)$ \\ 
        $\Omega_{cc}^{+}$ & $0.09(2)$ & $0.62(12)$ & $0.71(14)$ & $0$& $1.38(11)$ & $0.01$ & $0.07(1)$ & $1.45(12)$ \\ 
		$\Xi_{bb}^{0}$ & $0.60(12)$ & $3.72(74)$ & $4.31(86)$& $0$ & $9.74(175)$ & $0.06(1)$ & $0.41(8)$ & $10.22(246)$ \\ 
		$\Xi_{bb}^{-}$ & $0.50(1)$ & $3.03(61)$ & $3.53(71)$& $0$ & $1.83(50)$ & $0.05(1)$ & $0.34(7)$ & $2.23(71)$ \\ 
		$\Omega_{bb}^{-}$ & $0.09(2)$ & $0.69(14)$ & $0.79(16)$& $0$ & $1.26(34)$ & $0.01$ & $0.07(1)$ & $1.34(49)$ \\ 
		$\Xi_{b{\left\lbrace cu\right\rbrace }}^{+}$ & $5.00(100)$ & $0.96(19)$ & $5.97(119)$& $-1.97(40)$ & $4.37(56)$ & $0.46(9)$ & $0.11(2)$ & $2.96(97)$ \\ 
		$\Xi_{b{\left\lbrace cd\right\rbrace }}^{0}$ & $4.31(86)$ & $0.79(16)$ & $5.10(102)$& $-1.43(18)$ & $0.13(5)$ & $0.39(8)$ & $0.09(2)$ & $-0.83(25)$ \\ 
		$\Omega_{b{\left\lbrace cs\right\rbrace }}^{0}$ & $0.69(14)$ & $0.18(4)$ & $0.87(17)$& $-1.13(9)$ & $0.05(2)$ & $0.07(1)$ & $0.02$ & $-0.99(11)$ \\ 
		$\Xi_{b{\left[  cu\right]  }}^{+}$ &$1.36(27)$ & $2.32(46)$ & $3.68(74)$& $1.97(40)$ & $2.25(46)$ & $0.15(3)$ & $0.28(6)$ & $4.66(87)$ \\ 
		$\Xi_{b{\left[  cd\right]  }}^{0}$ & $1.11(22)$ & $1.84(37)$ & $2.94(59)$& $1.43(18)$ & $1.63(20)$ & $0.13(3)$ & $0.23(5)$ & $3.42(39)$ \\ 
		$\Omega_{b{\left[  cs\right]  }}^{0}$ & $0.26(5)$ & $0.48(10)$ & $0.74(15)$& $1.13(9)$ & $1.29(11)$ & $0.03(1)$ & $0.05(1)$ & $2.50(20)$ \\
         \hline
        & $\alpha_E^{(c'_1-g'_1)}$ & $\alpha_E^{(c'_2-g'_2)}$ & $\alpha_E^{\text{Tot.}}$& $\beta_M^{(b'_1)}$ & $\beta_M^{(b'_2)}$ & $\beta_M^{(c'_1-g'_1)}$ & $\beta_M^{(c'_2-g'_2)}$ &  $\beta_M^{\mathrm{Tot.}}$   \\ \hline
        $\Xi_{cc}^{*++}$ & $2.98(60)$ & $7.97(159)$ & $10.95(219)$& $0$ & $-1.20(25)$ & $0.30(6)$ & $0.42(8)$ & $-0.48(28)$\\ 
        $\Xi_{cc}^{*+}$ & $2.50(50)$ & $7.49(150)$ & $9.99(200)$& $0$ & $-0.87(11)$ & $0.25(5)$ & $0.38(8)$ & $-0.24(17)$ \\ 
        $\Omega_{cc}^{*+}$ & $0.47(9)$ & $0.48(10)$ & $0.96(19)$& $0$ & $-0.69(6)$ & $0.05(1)$ & $0.04(1)$ & $-0.60(6)$ \\ 
        $\Xi_{bb}^{*0}$ & $2.98(60)$ & $3.34(67)$ & $6.32(126)$& $0$ & $-4.87(87)$ & $0.30(6)$ & $0.28(6)$ & $-4.29(88)$\\ 
        $\Xi_{bb}^{*-}$ & $2.50(50)$ & $2.93(59)$ & $5.43(109)$& $0$ & $-0.92(25)$ & $0.25(5)$ & $0.24(5)$ & $-0.42(27)$ \\ 
        $\Omega_{bb}^{*-}$ & $0.47(9)$ & $0.42(8)$ & $0.89(18)$& $0$ & $-0.63(17)$ & $0.05(1)$ & $0.04(1)$ & $-0.54(17)$ \\ 
        $\Xi_{ bc}^{*+}$ & $2.98(60)$ & $4.86(97)$ & $7.84(157)$& $0$ & $-3.31(5)$ & $0.30(6)$ & $0.34(7)$ & $-2.68(14)$ \\ 
        $\Xi_{ bc}^{*0}$ & $2.50(50)$ & $4.41(88)$ & $6.92(138)$& $0$ & $-0.88(8)$ & $0.25(5)$ & $0.29(6)$ & $-0.33(13)$ \\ 
        $\Omega_{ bc}^{*0}$ & $0.47(9)$ & $0.22(4)$ & $0.69(14)$& $0$ & $-0.67(4)$ & $0.05(1)$ & $0.04(1)$ & $-0.58(5)$ \\
        \hline\hline
    \end{tabular}
\end{table}

The error estimates provided in Table~\ref{tab:polarizabilities_numerical_results_ccbaryon} account for the uncertainties in the input parameters, specifically the axial coupling constants ($g_i$, $c_i$) and the constituent quark magnetic moments $\mu_q$. The numerical results for the electromagnetic polarizabilities of doubly heavy baryons are shown in Table~\ref{tab:polarizabilities_numerical_results_ccbaryon}. For the baryons containing two identical heavy quarks ($ccq$ and $bbq$), the electric polarizabilities exhibit a pattern where the leading loop contributions are governed by the spin-changing intermediate states. Specifically, the spin-1/2 baryon $\mathcal{B}$ is predominantly influenced by the $\mathcal{B}^*\phi$ loops ($\alpha_E^{(c_2-g_2)}$), while its spin-3/2 counterpart $\mathcal{B}^*$ is mainly affected by the $\mathcal{B}\phi$ loops. This pattern contrasts with observations in singly heavy baryons and light-flavor baryons~\cite{Wen:2025xed,Chen:2024xks}. The discrepancy arises from differences in the spin–flavor structures of the different baryon multiplets, which result in distinct axial-vector couplings within the chiral loop diagrams.

In contrast, the $bcq$ system exhibits distinct behaviors due to the presence of the low-lying singlet state $\mathcal{T}$. For both the spin-1/2 state $\mathcal{B}$ and the spin-3/2 state $\mathcal{B}^*$, the electric polarizabilities are significantly enhanced by loop diagrams involving the $\mathcal{T}$ baryon. Consequently, the dominant contribution for $\mathcal{B}$ shifts to the spin-conserving channel $\alpha_E^{(c_1-g_1)}$ (which includes the $\mathcal{T}$ intermediate state), while for $\mathcal{B}^*$, it corresponds to the spin-changing channel $\alpha_E^{(c_2-g_2)}$ (which involves transitions to both $\mathcal{T}$ and $\mathcal{B}$).

Regarding the magnetic polarizabilities, the tree-level diagrams remain the dominant source for all doubly heavy baryons. However, a notable cancellation mechanism emerges for the $\mathcal{B}$ states in the $bcq$ sector. The magnetic transition to the lower-lying singlet $\mathcal{T}$ ($\beta_M^{(b_1)}$) yields a negative contribution, which destructively interferes with the positive contribution arising from the transition to the heavier doublet $\mathcal{B}^*$ ($\beta_M^{(b_2)}$). Consequently, the net magnetic polarizabilities of the $bcq$ baryons is determined by the competition between these two terms, resulting in values that can be either positive or negative.

Finally, it is instructive to examine the heavy quark limit ($m_Q \to \infty$), where the heavy quark symmetry becomes exact. In this limit, the mass splittings vanish ($\Delta, \delta \to 0$), and HDAS implies a dynamical equivalence between singly heavy mesons and doubly heavy baryons. To verify this explicitly, we recalculate the polarizabilities of singly heavy mesons using the axial coupling derived from the quark model ($g =0.6 g_N$) instead of the experimental extraction, ensuring consistency with the baryon sector. Remarkably, we find that the electromagnetic polarizabilities of singly heavy mesons and doubly heavy baryons with the same light quark become identical. Specifically, the electric polarizabilities are found to be $\alpha_E^{\mathrm{Tot.}} = 5.36$, $4.51$, and $0.85$ (in units of $10^{-4}$ fm$^3$) for the $u$, $d$, and $s$ light quarks, respectively, while the magnetic polarizabilities are exactly one-tenth of these values $\beta_M ^{\mathrm{Tot.}}= \alpha_E^{\mathrm{Tot.}} / 10$. Furthermore, the ratios of the coupling factors $A_{ch}$ listed in Table~\ref{tab:loop_params} strictly govern the relative weights of contributions from different intermediate loop channels. This exact matching serves as a robust consistency check of our calculation and confirms the restoration of HDAS in the heavy quark limit.

\section{Summary}\label{sec:DISCUSSION_AND_CONCULSION}
In this work, we have systematically calculated the electromagnetic polarizabilities of singly heavy mesons and doubly heavy baryons within the framework of HH$\chi$PT up to $\mathcal{O}(p^3)$. To derive quantitative predictions from the effective field theory, we have estimated the relevant LECs using the non-relativistic constituent quark model.

The most striking prediction of our study appears in the $D^*$ sector. We found that the electric polarizabilities of $\bar D^{*0}$ and $D^{*-}$ are enhanced by several orders of magnitude compared to their bottom counterparts or other heavy hadrons. This dramatic enhancement originates from a unique threshold mechanism: the mass splitting between $D^*$ and $D$ ($\Delta \approx 142$ MeV) is remarkably close to the charged pion mass ($m_\pi \approx 140$ MeV). This near-degeneracy leads to a threshold singularity (cusp-like structure) in the chiral loop functions. Consequently, the real part is strongly amplified by the long-range dynamics of the loosely bound pion cloud surrounding the $D^*$ meson, while the opening of the $\bar{D}^0 \pi^-$ decay channel generates a substantial imaginary part to the $D^{*-}$ polarizability.

For the doubly heavy baryons, the electromagnetic polarizabilities depend critically on the heavy-flavor composition. While polarizabilities of the $ccq$ and $bbq$ systems are dominated by spin-changing loops, the polarizabilities of the $bcq$ system exhibits distinct dynamics due to the presence of the low-lying scalar $S_{[cq]}=0$ state $\mathcal{T}$. Our analysis shows that the mixing with this singlet state fundamentally alters the polarization pattern: the electric polarizability of the $bcq$  axial-vector $S_{\{cq\}}=1$ state $\mathcal{B}$ becomes dominated by spin-conserving loops involving $\mathcal{T}$. Furthermore, the magnetic polarizability of this state is determined by a delicate cancellation, where the negative contribution from the transition to the lower-lying $\mathcal{T}$ destructively interferes with the positive contribution from the transition to the heavier $\mathcal{B}^*$.

Our results highlight the crucial role of chiral dynamics and kinematic thresholds in determining the electromagnetic structure of heavy hadrons. These derived expressions and numerical estimates provide necessary theoretical benchmarks for future lattice QCD simulations and experimental investigations.

\section*{ACKNOWLEDGMENTS}
This project was supported by the National
Natural Science Foundation of China (12475137). The computational resources were supported by the high-performance computing platform of Peking University.

\begin{appendix}

\section{Loop Integrals}\label{appendix:loop_integrals}
To combine propagator denominators, we introduce integrals over Feynman parameters:
\begin{equation}
	\frac{1}{A_1 A_2 \cdots A_n}=\int_0^1 d x_1 \cdots d x_n \delta\left(\sum x_i-1\right) \frac{(n-1)!}{\left[x_1 A_1+x_2 A_2+\cdots x_n A_n\right]^n}.
\end{equation}
To regularize divergent loop integrals, we use the dimensional regularization scheme and expand them around 4-dimensional spacetime. In this way, one can define the loop functions that frequently occur in calculations~~\cite{doi:10.1142/S0218301395000092,Scherer:2002tk,PhysRevD.55.5598}. Here, we list only those that we need:
\begin{equation}
\begin{aligned}
\frac{1}{i} &\int \frac{d^d \ell}{(2 \pi)^d} \frac{\left\{1, \ell_\mu \ell_\nu, \ell_\mu \ell_\nu \ell_\alpha \ell_\beta\right\}}{(v \cdot \ell-\omega-i \epsilon)\left(M_{\chi}^2-\ell^2-i \epsilon\right)}  \\
&=\left\{J_0\left(\omega, M_{\chi}^2\right)\right.,\quad g_{\mu \nu} J_2\left(\omega, M_{\chi}^2\right)+v_\mu v_\nu J_3\left(\omega, M_{\chi}^2\right),\quad \left.\left(g_{\mu \nu} g_{\alpha \beta}+\text { perm. }\right) J_6\left(\omega, M_{\chi}^2\right)+\ldots\right\}.
\end{aligned}
\end{equation}
All loop-integrals can be expressed via the basis-function $J_0$:
\begin{equation}\label{eq:J0_J2_J6}
\begin{aligned}
	J_0\left(\omega, M_{\chi}^2\right) & =-4 L \omega+\frac{\omega}{8 \pi^2}\left(1-2 \ln \frac{M_\chi}{\mu}\right)-\frac{1}{4 \pi^2} \sqrt{M_{\chi}^2-\omega^2} \arccos \frac{-\omega}{M_\chi}+\mathcal{O}(d-4),\\
J_2\left(\omega, M_{\chi}^2\right) & =\frac{1}{d-1}\left[\left(M_{\chi}^2-\omega^2\right) J_0\left(\omega, M_{\chi}^2\right)-\omega \Delta_{\chi}\right], \\
J_6\left(\omega, M_{\chi}^2\right) & =\frac{1}{d+1}\left[\left(M_{\chi}^2-\omega^2\right) J_2\left(\omega, M_{\chi}^2\right)-\frac{M_{\chi}^2 \omega}{d} \Delta_{\chi}\right].
\end{aligned}
\end{equation}
In Eq.~\eqref{eq:J0_J2_J6} we have used
\begin{equation}
\begin{aligned}
\Delta_{\chi} & =2 M_{\chi}^2\left(L+\frac{1}{16 \pi^2} \ln \frac{M_\chi}{\mu}\right) +\mathcal{O}(d-4),\\
L & =\frac{\mu^{d-4}}{16 \pi^2}\left[\frac{1}{d-4}+\frac{1}{2}\left(\gamma_E-1-\ln 4 \pi\right)\right],
\end{aligned}
\end{equation}
The $\gamma_E=0.557215$ is Euler constant. The scale $\mu$ is introduced in dimensional regularization.

For the spin-averaged forward Compton scattering amplitude, $J_i(-\omega-\delta)$ and $J_i(\omega-\delta)$ always appear symmetrically. Therefore, for simplicity, we define a new $\mathcal{J}$-function:
\begin{equation}
	\mathcal{J}_i(\omega,\delta,M_{\chi}^2)=J_i(\omega-\delta,M_{\chi}^2)+J_i(-\omega-\delta,M_{\chi}^2)
\end{equation}
With $\mathcal{J}_i^{\prime}$ and $\mathcal{J}_i^{\prime \prime}$ we define the first and second partial derivative with respect to $M_{\chi}^2$,

\begin{equation}
\begin{aligned}
\mathcal{J}_i^{\prime}\left(\omega,\delta ,M_{\chi}^2\right) & =\frac{\partial}{\partial\left(M_{\chi}^2\right)} \mathcal{J}_i\left(\omega, \delta,M_{\chi}^2\right) \\
\mathcal{J}_i^{\prime \prime}\left(\omega, \delta,M_{\chi}^2\right) & =\frac{\partial^2}{\partial\left(M_{\chi}^2\right)^2} \mathcal{J}_i\left(\omega,\delta, M_{\chi}^2\right)
\end{aligned}
\end{equation}
\section{The analytical expressions of electromagnetic polarizabilities}\label{appendix:expression_of_EMP}
Despite the distinct quark contents and spin configurations, the chiral loop diagrams for singly heavy mesons and doubly heavy baryons share identical topological structures determined by the chiral dynamics. Consequently, their contributions to the polarizabilities form factors can be described by a unified set of analytical expressions, parameterized by the channel-dependent coupling factors $A_{ch}$ and mass splittings $\delta_{ch}$. The non-vanishing form factors arising from the loop diagrams in Figs.~\ref{fig:spin0}($c$)--($g$) and Figs.~\ref{fig:spin1}($c$)--($g$) are expressed as follows:
\begin{align}
	U^{(c)}_{\xi}(\omega)= &\sum_{ch} A_{ch}\sum_{\phi}\frac{D^{(c)}_{\xi,\phi}}{F_\phi^2}\mathcal{J}_0(\omega,\delta_{ch},M_{\phi}^2),\\
	U^{(d+e)}_{\xi}(\omega)= & \sum_{ch} A_{ch}\left[\sum_{\phi} \frac{D_{\xi,\phi}^{(d+e)}}{F_\phi^2}\int_0^1 dx~\mathcal{J}^{\prime}_2(\omega x,\delta_{ch},M_{\phi}^2)\right],\\
	U^{(f)}_{\xi}(\omega)= &\sum_{ch} A_{ch}\left[\sum_{\phi} \frac{D_{\xi,\phi}^{(f)}}{F_\phi^2}\int_0^1dx~ (1-x)(d+1)\mathcal{J}_6^{\prime\prime}(\omega x,\delta_{ch},M_{\phi}^2)\right.\notag \\
	&\quad ~ -\left. \sum_{\phi} \frac{D_{\xi,\phi}^{(f)}}{F_\phi^2} \int_0^1dx~\omega^2(1-x)x^2\mathcal{J}_2^{\prime\prime}(\omega x,\delta_{ch},M_{\phi}^2)\right],\\
	U^{(g)}_{\xi}(\omega)= &\sum_{ch} A_{ch}\sum_{\phi}\frac{D^{(g)}_{\xi,\phi}}{F_\phi^2}(d-1)\mathcal{J}_2^{\prime}(0,\delta_{ch},M_{\phi}^2),\\
	V^{(c)}_{\xi}(\omega)= & 0,\\
	V^{(d+e)}_{\xi}(\omega)= & \sum_{ch} A_{ch}\left[\sum_{\phi} -\frac{1}{2 F_\phi^2}D_{\xi,\phi}^{(d+e)}\int_0^1 dx~x(1-2x)\mathcal{J}^{\prime}_0(\omega x,\delta_{ch},M_{\phi}^2)\right],\\
	V^{(f)}_{\xi}(\omega)= & \sum_{ch} A_{ch}\left[\sum_{\phi} \frac{1}{4 F_\phi^2}D^{(f)}_{\xi,\phi}\int_0^1dx~(1-x)\left[8x(2x-1)+(2x-1)^2(d-1)\right]\mathcal{J}_2^{\prime\prime}(\omega x,\delta_{ch},M_{\phi}^2)\right.\notag \\
	 &\quad ~ -\left. \sum_{\phi} \frac{1}{4 F_\phi^2}D^{(f)}_{\xi,\phi} \int_0^1 dx~ \omega^2(1-x)x^2(2x-1)^2 \mathcal{J}_0^{\prime\prime}(\omega x,\delta_{ch},M_{\phi}^2)\right],\\
	 V^{(g)}_{\xi}(\omega)= & 0,
\end{align}
Here, the summation $\sum_{ch}$ runs over all allowed intermediate loop channels for a given external state $\xi$. The kinematic coefficients $D_{\xi,\phi}$ depend only on the flavor representation and are listed in Table~\ref{tab:loop_cofficients}. The channel-specific parameters $A_{ch}$ and $\delta_{ch}$ are summarized in Table~\ref{tab:loop_params}

\begin{table*}[htbp]
\renewcommand{\arraystretch}{1.5}
\caption{The coupling factors $A_{ch}$ and mass splittings $\delta_{ch}$ for the loop diagrams. The summation in the analytical expressions runs over the intermediate loop channels listed below for each external state $\xi$.}
\label{tab:loop_params}
\begin{tabular}{l|c|c|c}
\hline\hline
External State ($\xi$) & Intermediate Loop Channel & Coupling Factor ($A_{ch}$) & Mass Splitting ($\delta_{ch}$) \\ \hline
\multicolumn{4}{l}{\textbf{Singly Heavy Mesons}} \\ \hline
$P$ ($0^-$) & $P \phi$ & $0$ & $0$ \\ 
& $P^* \phi$ & $\frac{1}{2}e^2g^2$ & $\Delta$ \\ \hline
$P^*$ ($1^-$) & $P \phi$ & $\frac{1}{6}e^2g^2$ & $-\Delta$ \\
& $P^* \phi$ & $\frac{1}{3}e^2g^2$ & $0$ \\ \hline
\multicolumn{4}{l}{\textbf{Doubly Heavy Baryons ($ccq/bbq$)}} \\ \hline
$\mathcal{B}$ ($1/2^+$) & $\mathcal{B} \phi$ & $\frac{1}{2}e^2g_1^2$ & $0$ \\ 
& $\mathcal{B}^* \phi$ & $\frac{1}{2}e^2g_2^2 \frac{d-2}{d-1}$ & $\delta$ \\ \hline
$\mathcal{B}^*$ ($3/2^+$) & $\mathcal{B} \phi$ & $\frac{1}{4}e^2g_2^2$ & $-\delta$ \\
& $\mathcal{B}^* \phi$ & $\frac{d^3-4d^2+d+6}{4(d-1)^2}e^2g_3^2$ & $0$ \\ \hline
\multicolumn{4}{l}{\textbf{Doubly Heavy Baryons ($bcq$)}} \\ \hline
$\mathcal{T}$ ($1/2^+$) & $\mathcal{T} \phi$ & $\frac{1}{2}e^2c_1^2$ & $0$ \\
& $\mathcal{B} \phi$ & $\frac{1}{2}e^2c_2^2$ & $\delta_1$ \\
& $\mathcal{B}^* \phi$ & $\frac{1}{2}e^2c_3^2 \frac{d-2}{d-1}$ & $\delta_2$ \\ \hline
$\mathcal{B}$ ($1/2^+$) & $\mathcal{B} \phi$ & $\frac{1}{2}e^2g_1^2$ & $0$ \\
& $\mathcal{T} \phi$ & $\frac{1}{2}e^2c_2^2$ & $-\delta_1$ \\
& $\mathcal{B}^* \phi$ & $\frac{1}{2}e^2g_2^2 \frac{d-2}{d-1}$ & $\delta_2-\delta_1$ \\ \hline
$\mathcal{B}^*$ ($3/2^+$) & $\mathcal{T} \phi$ & $\frac{1}{4}e^2c_3^2$ & $-\delta_2$ \\
& $\mathcal{B} \phi$ & $\frac{1}{4}e^2g_2^2$ & $\delta_1-\delta_2$ \\
& $\mathcal{B}^* \phi$ & $\frac{d^3-4d^2+d+6}{4(d-1)^2}e^2g_3^2$ & $0$ \\ \hline\hline
\end{tabular}
\end{table*}

\begin{table*}[htbp]
\renewcommand{\arraystretch}{1.5}
\setlength{\tabcolsep}{8pt}
\caption{The coefficients $D_{\xi,\phi}$ for the loop diagrams. These coefficients depend solely on the light quark flavor $q$ of the external hadron $\xi$. The specific hadrons corresponding to each light quark flavor are: $q=u$ ($\bar{D}^0, \Xi_{cc}^{++}$, ...), $q=d$ ($D^-, \Xi_{cc}^+$, ...), and $q=s$ ($D_s^-, \Omega_{cc}^+$, ...).}
\label{tab:loop_cofficients}
\begin{tabular}{l|cccc|cccc}
\hline\hline
\multirow{2}{*}{Light Quark ($q$)} & \multicolumn{4}{c|}{$\pi$-loop coefficients ($D_{\xi,\pi}$)} & \multicolumn{4}{c}{$K$-loop coefficients ($D_{\xi,K}$)} \\ \cline{2-9}
 & $(c)$ & $(d+e)$ & $(f)$ & $(g)$ & $(c)$ & $(d+e)$ & $(f)$ & $(g)$ \\ \hline
$u$ & $-1$ & $4$ & $-4$ & $1$ & $1$ & $4$ & $-4$ & $1$ \\
$d$ & $-1$ & $4$ & $-4$ & $1$ & $0$ & $0$ & $0$ & $0$ \\
$s$ & $0$ & $0$ & $0$ & $0$ & $-1$ & $4$ & $-4$ & $1$ \\ \hline\hline
\end{tabular}
\end{table*}

For convenience, we just list the results for charmed hadrons:
\begin{align}
    \alpha_E^{(c_1-g_1)}\left(\bar{D}^0\right)=&\alpha_E^{(c_1-g_1)}\left(D^-\right)=\alpha_E^{(c_1-g_1)}\left(D_s^-\right)=0,\\
    \alpha_E^{(c_2-g_2)}\left(\bar{D}^0\right)=&\frac{\alpha_{\mathrm{em}} g^2 S_\pi(\Delta)}{96\pi^2 F_\pi^2 \left(M_\pi^2-\Delta^2\right)^2}+\frac{\alpha_{\mathrm{em}} g^2 S_K(\Delta)}{96\pi^2 F_K^2 \left(M_K^2-\Delta^2\right)^2},\\
    \alpha_E^{(c_2-g_2)}\left(D^-\right)=&\frac{\alpha_{\mathrm{em}} g^2 S_\pi(\Delta)}{96\pi^2 F_\pi^2 \left(M_\pi^2-\Delta^2\right)^2},\\
    \alpha_E^{(c_2-g_2)}\left(D_s^-\right)=&\frac{\alpha_{\mathrm{em}} g^2 S_K(\Delta)}{96\pi^2 F_K^2 \left(M_K^2-\Delta^2\right)^2},\\
    \alpha_E^{(c_1-g_1)}\left(\Xi_{cc}^{++}\right)=&\frac{5\alpha_{\mathrm{em}} g_1^2 }{96\pi F_\pi^2 M_\pi}+\frac{5\alpha_{\mathrm{em}} g_1^2 }{96\pi F_K^2 M_K},\\
    \alpha_E^{(c_1-g_1)}\left(\Xi_{cc}^{+}\right)=&\frac{5\alpha_{\mathrm{em}} g_1^2 }{96\pi F_\pi^2 M_\pi},\\
    \alpha_E^{(c_1-g_1)}\left(\Omega_{cc}^{+}\right)=&\frac{5\alpha_{\mathrm{em}} g_1^2 }{96\pi F_K^2 M_K},\\
    \alpha_E^{(c_2-g_2)}\left(\Xi_{cc}^{++}\right)=&\frac{\alpha_{\mathrm{em}} g_2^2 S_\pi(\delta)}{144\pi^2 F_\pi^2 \left(M_\pi^2-\delta^2\right)^2}+\frac{\alpha_{\mathrm{em}} g_2^2 S_K(\delta)}{144\pi^2 F_K^2 \left(M_K^2-\delta^2\right)^2},\\
    \alpha_E^{(c_2-g_2)}\left(\Xi_{cc}^{+}\right)=&\frac{\alpha_{\mathrm{em}} g_2^2 S_\pi(\delta)}{144\pi^2 F_\pi^2 \left(M_\pi^2-\delta^2\right)^2},\\
    \alpha_E^{(c_2-g_2)}\left(\Omega_{cc}^{+}\right)=&\frac{\alpha_{\mathrm{em}} g_2^2 S_K(\delta)}{144\pi^2 F_K^2 \left(M_K^2-\delta^2\right)^2},\\
    \alpha_E^{(c^\prime_1-g^\prime_1)}\left(\bar{D}^{0*}\right)=&\frac{5\alpha_{\mathrm{em}} g^2 }{144\pi F_\pi^2 M_\pi}+\frac{5\alpha_{\mathrm{em}} g^2 }{144\pi F_K^2 M_K},\\
    \alpha_E^{(c^\prime_1-g^\prime_1)}\left(D^{-*}\right)=&\frac{5\alpha_{\mathrm{em}} g^2 }{144\pi F_\pi^2 M_\pi},\\
    \alpha_E^{(c^\prime_1-g^\prime_1)}\left(D_s^{-*}\right)=&\frac{5\alpha_{\mathrm{em}} g^2 }{144\pi F_K^2 M_K},\\
    \alpha_E^{(c^\prime_2-g^\prime_2)}\left(\bar{D}^{0*}\right)=&\frac{\alpha_{\mathrm{em}} g^2 S_\pi(-\Delta)}{288\pi^2 F_\pi^2 \left(M_\pi^2-\Delta^2\right)^2}+\frac{\alpha_{\mathrm{em}} g^2 S_K(-\Delta)}{288\pi^2 F_K^2 \left(M_K^2-\Delta^2\right)^2},\\
    \alpha_E^{(c^\prime_2-g^\prime_2)}\left(D^{-*}\right)=&\frac{\alpha_{\mathrm{em}} g^2 S_\pi(-\Delta)}{288\pi^2 F_\pi^2 \left(M_\pi^2-\Delta^2\right)^2},\\
    \alpha_E^{(c^\prime_2-g^\prime_2)}\left(D_s^{-*}\right)=&\frac{\alpha_{\mathrm{em}} g^2 S_K(-\Delta)}{288\pi^2 F_K^2 \left(M_K^2-\Delta^2\right)^2},\\
    \alpha_E^{(c^\prime_1-g^\prime_1)}\left(\Xi_{cc}^{++*}\right)=&\frac{25\alpha_{\mathrm{em}} g_3^2 }{864\pi F_\pi^2 M_\pi}+\frac{25\alpha_{\mathrm{em}} g_3^2 }{864\pi F_K^2 M_K},\\
    \alpha_E^{(c^\prime_1-g^\prime_1)}\left(\Xi_{cc}^{+*}\right)=&\frac{25\alpha_{\mathrm{em}} g_3^2 }{864\pi F_\pi^2 M_\pi},\\
    \alpha_E^{(c^\prime_1-g^\prime_1)}\left(\Omega_{cc}^{+*}\right)=&\frac{25\alpha_{\mathrm{em}} g_3^2 }{864\pi F_K^2 M_K},\\
    \alpha_E^{(c^\prime_2-g^\prime_2)}\left(\Xi_{cc}^{++*}\right)=&\frac{\alpha_{\mathrm{em}} g_2^2 S_\pi(-\delta)}{288\pi^2 F_\pi^2 \left(M_\pi^2-\delta^2\right)^2}+\frac{\alpha_{\mathrm{em}} g_2^2 S_K(-\delta)}{288\pi^2 F_K^2 \left(M_K^2-\delta^2\right)^2},\\
    \alpha_E^{(c^\prime_2-g^\prime_2)}\left(\Xi_{cc}^{+*}\right)=&\frac{\alpha_{\mathrm{em}} g_2^2 S_\pi(-\delta)}{288\pi^2 F_\pi^2 \left(M_\pi^2-\delta^2\right)^2},\\
    \alpha_E^{(c^\prime_2-g^\prime_2)}\left(\Omega_{cc}^{+*}\right)=&\frac{\alpha_{\mathrm{em}} g_2^2 S_K(-\delta)}{288\pi^2 F_K^2 \left(M_K^2-\delta^2\right)^2},
\end{align}
\begin{align}
    \beta_M^{(c_1-g_1)}\left(\bar{D}^0\right)=&\beta_M^{(c_1-g_1)}\left(D^-\right)=\beta_M^{(c_1-g_1)}\left(D_s^-\right)=0,\\
    \beta_M^{(c_2-g_2)}\left(\bar{D}^0\right)=&\frac{\alpha_{\mathrm{em}} g^2 R_\pi(\Delta)}{96\pi^2 F_\pi^2 \left(M_\pi^2-\Delta^2\right)}+\frac{\alpha_{\mathrm{em}} g^2 R_K(\Delta)}{96\pi^2 F_K^2 \left(M_K^2-\Delta^2\right)},\\
    \beta_M^{(c_2-g_2)}\left(D^-\right)=&\frac{\alpha_{\mathrm{em}} g^2 R_\pi(\Delta)}{96\pi^2 F_\pi^2 \left(M_\pi^2-\Delta^2\right)},\\
    \beta_M^{(c_2-g_2)}\left(D^-_s\right)=&\frac{\alpha_{\mathrm{em}} g^2 R_K(\Delta)}{96\pi^2 F_K^2 \left(M_K^2-\Delta^2\right)},\\
    \beta_M^{(c_1-g_1)}\left(\Xi_{cc}^{++}\right)=&\frac{\alpha_{\mathrm{em}} g_1^2 }{192\pi F_\pi^2 M_\pi}+\frac{\alpha_{\mathrm{em}} g_1^2 }{192\pi F_K^2 M_K},\\
    \beta_M^{(c_1-g_1)}\left(\Xi_{cc}^{+}\right)=&\frac{\alpha_{\mathrm{em}} g_1^2 }{192\pi F_\pi^2 M_\pi},\\
    \beta_M^{(c_1-g_1)}\left(\Omega_{cc}^{+}\right)=&\frac{\alpha_{\mathrm{em}} g_1^2 }{192\pi F_K^2 M_K},\\
    \beta_M^{(c_2-g_2)}\left(\Xi_{cc}^{++}\right)=&\frac{\alpha_{\mathrm{em}} g_2^2 R_\pi(\delta)}{144\pi^2 F_\pi^2 \left(M_\pi^2-\delta^2\right)}+\frac{\alpha_{\mathrm{em}} g_2^2 R_K(\delta)}{144\pi^2 F_K^2 \left(M_K^2-\delta^2\right)},\\
    \beta_M^{(c_2-g_2)}\left(\Xi_{cc}^{+}\right)=&\frac{\alpha_{\mathrm{em}} g_2^2 R_\pi(\delta)}{144\pi^2 F_\pi^2 \left(M_\pi^2-\delta^2\right)},\\
    \beta_M^{(c_2-g_2)}\left(\Omega_{cc}^{+}\right)=&\frac{\alpha_{\mathrm{em}} g_2^2 R_K(\delta)}{144\pi^2 F_K^2 \left(M_K^2-\delta^2\right)},\\
    \beta_M^{(c^\prime_1-g^\prime_1)}\left(\bar{D}^{0*}\right)=&\frac{\alpha_{\mathrm{em}} g^2 }{288\pi F_\pi^2 M_\pi}+\frac{\alpha_{\mathrm{em}} g^2 }{288\pi F_K^2 M_K},\\
    \beta_M^{(c^\prime_1-g^\prime_1)}\left(D^{-*}\right)=&\frac{\alpha_{\mathrm{em}} g^2 }{288\pi F_\pi^2 M_\pi},\\
    \beta_M^{(c^\prime_1-g^\prime_1)}\left(D_s^{-*}\right)=&\frac{\alpha_{\mathrm{em}} g^2 }{288\pi F_K^2 M_K},\\
    \beta_M^{(c^\prime_2-g^\prime_2)}\left(\bar{D}^{0*}\right)=&\frac{\alpha_{\mathrm{em}} g^2 R_\pi(-\Delta)}{288\pi^2 F_\pi^2 \left(M_\pi^2-\Delta^2\right)}+\frac{\alpha_{\mathrm{em}} g^2 R_K(-\Delta)}{288\pi^2 F_K^2 \left(M_K^2-\Delta^2\right)},\\
    \beta_M^{(c^\prime_2-g^\prime_2)}\left(D^{-*}\right)=&\frac{\alpha_{\mathrm{em}} g^2 R_\pi(-\Delta)}{288\pi^2 F_\pi^2 \left(M_\pi^2-\Delta^2\right)},\\
    \beta_M^{(c^\prime_2-g^\prime_2)}\left(D^{-*}_s\right)=&\frac{\alpha_{\mathrm{em}} g^2 R_K(-\Delta)}{288\pi^2 F_K^2 \left(M_K^2-\Delta^2\right)},\\
     \beta_M^{(c^\prime_1-g^\prime_1)}\left(\Xi_{cc}^{++*}\right)=&\frac{5\alpha_{\mathrm{em}} g_3^2 }{1728\pi F_\pi^2 M_\pi}+\frac{5\alpha_{\mathrm{em}} g_3^2 }{1728\pi F_K^2 M_K},\\
    \beta_M^{(c^\prime_1-g^\prime_1)}\left(\Xi_{cc}^{+*}\right)=&\frac{5\alpha_{\mathrm{em}} g_3^2 }{1728\pi F_\pi^2 M_\pi},\\
    \beta_M^{(c^\prime_1-g^\prime_1)}\left(\Omega_{cc}^{+*}\right)=&\frac{5\alpha_{\mathrm{em}} g_3^2 }{1728\pi F_K^2 M_K},\\
    \beta_M^{(c^\prime_2-g^\prime_2)}\left(\Xi_{cc}^{++*}\right)=&\frac{\alpha_{\mathrm{em}} g_2^2 R_\pi(-\delta)}{288\pi^2 F_\pi^2 \left(M_\pi^2-\delta^2\right)}+\frac{\alpha_{\mathrm{em}} g_2^2 R_K(-\delta)}{288\pi^2 F_K^2 \left(M_K^2-\delta^2\right)},\\
    \beta_M^{(c^\prime_2-g^\prime_2)}\left(\Xi_{cc}^{+*}\right)=&\frac{\alpha_{\mathrm{em}} g_2^2 R_\pi(-\delta)}{288\pi^2 F_\pi^2 \left(M_\pi^2-\delta^2\right)},\\
    \beta_M^{(c^\prime_2-g^\prime_2)}\left(\Omega_{cc}^{+*}\right)=&\frac{\alpha_{\mathrm{em}} g_2^2 R_K(-\delta)}{288\pi^2 F_K^2 \left(M_K^2-\delta^2\right)},
\end{align}
where we have defined
\begin{equation}\begin{aligned}
 & R_{\chi}(\delta)=\sqrt{M_\chi^2-\delta^2}\arccos\left[\frac{\delta}{M_\chi}\right], \\
 & S_{\chi}(\delta)=M_{\chi}^2\left(10R_{\chi}(\delta)-9\delta\right)+\delta^2\left(9\delta-R_{\chi}(\delta)\right).
\end{aligned}\end{equation}
By summing all the contributions calculated above, we can obtain the total electromagnetic polarizabilities:
\begin{align}
    \alpha_E^{\mathrm{Tot.}}(\xi) &=\sum_{i=1,2}\alpha_E^{(c_i-g_i)}(\xi),\quad
    \alpha_E^{\mathrm{Tot.}}(\xi^*) =\sum_{i=1,2}\alpha_E^{(c^\prime_i-g^\prime_i)}(\xi^*),\\
	\beta_M^{\mathrm{Tot.}}(\xi) &=\beta_M^{(b_2)}(\xi)+\sum_{i=1,2}\beta_M^{(c_i-g_i)}(\xi), \quad \beta_M^{\mathrm{Tot.}}(\xi^*) =\beta_M^{(b^\prime_2)}(\xi^*)+\sum_{i=1,2}\beta_M^{(c_i^\prime-g_i^\prime)}(\xi^*).
\end{align}

\end{appendix}
\bibliography{references}
\end{document}